# FLEXO-INDUCED FERROELECTRICITY IN LOW DIMENSIONAL TRANSITION METAL DICHALCOGENIDES

Anna N. Morozovska[1*], Eugene A. Eliseev[2], Galina I. Dovbeshko[1], Maya D. Glinchuk[2], Yunseok Kim[3†], and Sergei V. Kalinin[4,‡]

[1]*Institute of Physics, National Academy of Sciences of Ukraine,*
*46, pr. Nauky, 03028 Kyiv, Ukraine*

[2] *Institute for Problems of Materials Science, National Academy of Sciences of Ukraine, Krjijanovskogo 3, 03142 Kyiv, Ukraine*

[3] *School of Advanced Materials Science and Engineering, Sungkyunkwan University (SKKU), Suwon 16419, Republic of Korea*

[4]*The Center for Nanophase Materials Sciences, Oak Ridge National Laboratory,*
*Oak Ridge, TN 37922*

**Abstract**

We developed a Landau type theory for the description of polar phenomena in low-dimensional transition metal dichalcogenides (TMDs), specifically exploring flexoelectric origin of the polarization induced by a spontaneous bending and by inversion symmetry breaking due to the interactions with substrate. We consider the appearance of the spontaneous out-of-plane polarization due to the flexoelectric coupling with the strain gradient of the spontaneous surface rippling and surface-induced piezoelectricity. Performed calculations proved that the out-of-plane spontaneous polarization, originated from flexoelectric effect in a rippled TMD, is bistable and reversible by a non-uniform electric field. In contrast, the spontaneous polarization induced by a misfit strain and symmetry-sensitive surface-induced piezoelectric coupling, cannot be reversed by an external electric field. The special attention is paid to the spectral analysis of the linear dielectric susceptibility and gain factor, which enhancement is critically important for the observation of the polar phenomena in low-dimensional TMDs by the surface-enhanced vibrational spectroscopy.

---

[*] Corresponding author 1: anna.n.morozovska@gmail.com

[†] Corresponding author 2: yunseokkim@skku.edu

[‡] Corresponding author 3: sergei2@ornl.gov

# I. INTRODUCTION

Polar materials with ferroelectric and piezoelectric properties have been extensively investigated due to numerous device applications such as random-access memories, field effect transistors, and actuators [1, 2]. While ferroelectricity in polar materials originates from multiple intrinsic origins, such as temperature-induced spontaneous symmetry breaking, geometric distortion and charge ordering; it was also predicted theoretically that polar nature at the nanoscale can originate from extrinsic or intrinsic strain gradients due to the flexoelectricity, and surface-induced piezoelectric effects [3, 4, 5, 6]. However, until recently it was nearly impossible to experimentally probe flexoelectric or surface-induced piezoelectric effects because of negligibly small contribution of surfaces in the functionalities of bulk materials.

Recent emergence of low dimensional transition metal dichalcogenides (**TMDs**) opened a new pathway to explore polar phenomena originated from flexoelectric and surface-induced piezoelectric effects by providing model atomically thin environmentally stable materials. The polar phenomena in low dimensional TMDs such as $MoS_2$, $ReS_2$, and $MoTe_2$, that are centrosymmetric in a bulk form [7], can have multiple origins, including surface-induced phase transitions, local strain, and bending [8, 9]. In particular, it was predicted theoretically, that the centrosymmetric TMDs, such as $MoTe_2$ and $MoS_2$, can become polar in a single layer (**SL**) form [8, 9]. These predictions were ascertained by the experimental observations [7], where the emergence of odd-even parity of electromechanical response with number of layers [10] was interpreted as a clear evidence for the polar phenomena. Furthermore, recent works shows that SL $MoS_2$ [11] and $MoTe_2$ [12] can even exhibit ferroelectric properties due to the spontaneous symmetry breaking for low dimensions. Similarly, the multiple polar domain types were shown to exist in different structural phases of $MoS_2$, $WSe_2$, and $MoTe_2$. These can be metallic, semi-metallic or semi-conducting [13, 14, 15, 16] dependent on the local strains, corrugations, doping and related factors [11, 17, 18, 19, 20]. Another mechanism of the polar phenomena in the SL TMDs is the inversion symmetry breaking at the TMD surface completely or partially clamped to a substrate. In this case, polarization can originate from the surface-induced piezoelectric effect [21, 22].

Note, that some TMDs, such as $WTe_2$, are non-centrosymmetric in a bulk form [23]. Interestingly, for these materials, single layers can be either centrosymmetric [23], or non-centrosymmetric [8, 9]. For instance, two- or three- layered $WTe_2$ was found to exhibit switchable out-of-plane ferroelectricity [24], and the polar axis orientation correlates with electronic anisotropy in layered materials as similar to results [25]. However, we still need comprehensive framework to understand origin of ferroelectricity in low dimensional TMD based on its key aspect of physical properties.

The key aspect of the physics of SL-TMDs is that these materials can be readily bent when suspended. The bending is not limited to the micrometer-scale suspended films, but can also manifest as atomic-scale rippling [26], which is an effective pathway to reduce in-plane strain, or nanometer-scale surface corrugation correlated with substrate roughness. As predicted by several authors [27, 28], and later discovered experimentally, new degrees of freedom due to this bending can induce polar phenomena, either rintrinsic or induced by the external structural and charge disorder. Large curvatures enabled by small bending stiffnesses can give rise to significant flexoelectric effect induced polarization [29, 30, 31] The bending-induced out-of-plane dipole moment with density $p \sim (0.01 - 0.4)$ C/nm and flexoelectric polarization $P \sim (1 - 2) \times 10^{-3}$ C/m$^2$ were calculated from the first principles for MoS$_2$ [18 - 20], WTe$_2$ [8, 9] and WS$_2$ [31], respectively. Thus, the bending can play a significant role in the polar physics of 2D TMDs.

However, new degrees of freedom due to bending in SL-TMDs and its role on the ferroelectricity are still relatively unexplored. In particular, while the origin of the out-of-plane polarization in SL-TMDs is manifold, it can be the spontaneous surface reconstruction to the "rippled" state with the lowest energy, and/or corrugation, (i.e. bending-induced "flexo-origin") joined with the inversion symmetry breaking at the surface (e.g. surface-induced "piezo-origin"). Here, we develop a general mesoscopic phenomenological theory for polar phenomena in low dimensional TMDs that allows for analysis out-of-plane polarization instability induced by the bending-related flexoelectricity and surface-induced inversion symmetry breaking. Our aim is to explore the emergence of the ferroelectric-like phenomena in the low-dimensional systems described by a thermodynamic potential in response to the strain gradient of the reconstructed surface, which rippling is the lowest energy state in agreement with DFT studies [18 - 20, 31].

**B. Qualitative Mechanisms for Polarization Emergence**

In the both bending-induced flexo-origin and surface-induced piezo-origin, the polarization is spontaneous, and the overall picture for the flexo- and piezo- mechanisms of polarization origin is shown schematically in **Fig. 1.** Here, **Fig. 1a** illustrates the cross-section of an "unreconstructed" flat SL-TMD with a chemical formula MX$_2$, where M is a transition metal, X is a chalcogenide. Here M ions projection is located at the middle line between X ions projection. The strain is absent in a flat SL. Since the effective (Born or Bader) charges $Q$ of M and X ions are opposite, namely $Q_M = -2Q_X$, the total polarization is absent.

A mechanically free (i.e. "freestanding") SL-MX$_2$, where all X and M ions positions can spontaneously reconstruct to the rippled state with the lowest energy, is shown in **Fig. 1b**. Since the force matrix is different for the "light" X and "heavy" M ions, the amplitudes of X and M displacements are different for the periodic ripples. Since the effective charges are opposite, $Q_M = -2Q_X$, the periodic displacement of X and M ions induces the out-of-plane polarization modulation (dashed black curve and arrows) due to the flexoelectric coupling. The average polarization is zero in the case, and the periodic modulation of its out-of-plane component $P_3$ is proportional to the second derivative of elastic displacement and flexoelectric coefficient, namely the $P_3 \cong f_{3jkl} \frac{\partial^2 U_j}{\partial x_k \partial x_l} - \mu_{3j} \frac{\partial^2 U_j}{\partial t^2}$. Here $U$ is the mechanical displacement, $f_{ijkl}$ is the static flexoelectric tensor [32] determined by the microscopic properties of the material [33, 34], $\mu_{ij}$ is the dynamic flexoelectric tensor [35]. Also, a purely electronic flexoelectric polarization can be induced by bending of graphene-like semiconductors [27, 27].

A rippled SL-MX$_2$, which bottom X ions are clamped to a rigid substrate, is shown in **Fig. 1c**. The misfit strain $u_m$ appears from the MX$_2$ and substrate lattice constants mismatch. Since the bottom X ions are bonded to the substrate [36], they remained "clamped", and the small reconstruction is possible for the middle M ions, and a little bigger reconstruction for the top X ions. The small harmonic-like periodic displacement of the middle M and top X ions induces the out-of-plane polarization (dashed black curve and arrows) due to the flexoelectric coupling and surface-induced piezoelectric effect. The surface-induced piezoelectric effect is a result of the inversion symmetry breaking [21, 22], manifesting itself in different displacements of the top and bottom X ions from the middle line – M ions due to the interaction with substrate. The emerging out-of-plane polarization can be estimated as $P_3 \cong f_{3jkl} \frac{\partial^2 U_j}{\partial x_k \partial x_l} - \mu_{3j} \frac{\partial^2 U_j}{\partial t^2} + d_{31}^S u_m$, where the first two terms have the flexoelectric origin, and the third term has the piezoelectric origin, and $d_{ij}^S$ is the tensor of the surface-induced piezoelectric effect. The flexoelectric contribution exists for all possible symmetries [3 - 32] of the SL, while the piezoelectric one is symmetry sensitive [21, 22].

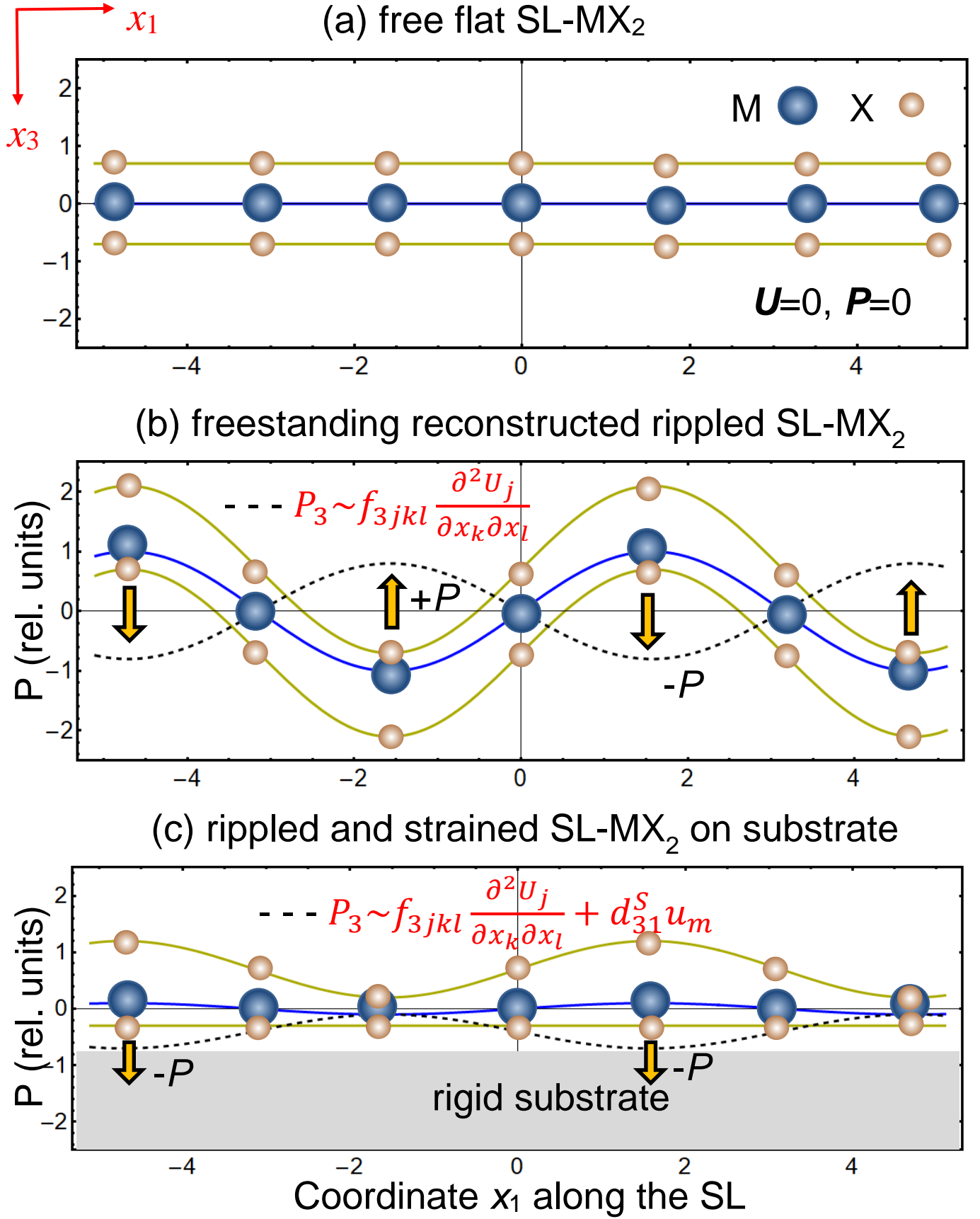

**FIGURE 1. Bending-induced out-of-plane polarization in a SL-TMD with a chemical formula MX$_2$.** M is a transition metal, X is a chalcogenide. M ions are positively charged, and X ions are negatively charged. **(a)** For flat freestanding SL-MX$_2$, strain $U$ and polarization $P$ are absent. **(b)** A freestanding SL-MX$_2$, where all X and M ions can spontaneously relax to the rippled state of the lowest energy. Since the force matrix is different for the "light" X atoms

and "heavy" M ions, the rippling results into the different amplitude of X and M ionic displacements. The harmonic-like periodic displacement of X and M ions induces the spatially modulated out-of-plane polarization (dashed black curve and arrows) due to the flexoelectric coupling. Orange arrows indicate the direction of emerging dipole polarization P. **(c)** Rippled and strained SL-MX$_2$, clamped to a rigid substrate. The misfit strain $u_m$ appears from the lattice constants mismatch of MX$_2$ and substrate. Here the bottom X ions are clamped to the substrate and remained at place, the small change of position is possible for the middle M ions, and a little bigger reconstruction occurs for the top surface X atoms. The small harmonic-like periodic displacement of the middle M ions and top X ions induces the polarization (dashed black curve and arrows) due to the surface-induced piezoelectric effect and flexoelectric coupling.

The emergence of the spontaneous polarization [shown in **Figs.1b-c**] is a linear response of the SL-TMD to the periodic static bending of the layer, i.e. the picture is a static "snapshot". The static polarization is spontaneous, since the flexo-induced and piezo-induced contributions to the strain and its gradient are spontaneous. However, the bistable polarization states, required for the appearance of robust ferroelectricity, cannot be described within the linear model and requires considering possible non-linear mechanisms. One of the suitable mechanisms can be the paraelectric nonlinearity, responsible for the appearance of a strain-induced (in fact ferroelastic) ferroelectricity in such materials as ultra-thin films of SrTiO$_3$ [37], TiO$_2$ [38] and HfO$_2$ [39, 40]. The nonlinearity is added as the small positive term $bP^4$ to the polarization part of the free energy, $F[P] = aP^2 + bP^4 + quP^2 + \cdots$, where $a$ and $b>0$ are expansion coefficients on polarization powers, $q$ is an electrostriction coefficient, and $u$ is the strain component, respectively. While the linear dielectric susceptibility in the free energy is positive ($a>0$), the nonlinear term is insignificant, but if the strain effects can change the sign of the linear term (i.e. $a^*<0$), the nonlinearity can become responsible for the appearance of bi-stable (or multi-stable) ferroelectric-like polarization hysteresis.

The formulation of a phenomenological Landau-type model, with a special attention to the free energy functional, local electric fields and polarization dynamics is given in **Section II**. The properties of the strain-induced polarization, linear dielectric susceptibility and nonlinear polarization response to external electric field are analyzed in **Section III**. **Section IV** is devoted the analysis of the gain factor, which enhancement is critically important for the observation of the ferroelectric-like phenomena in low dimensional TMDs by the surface-enhanced vibrational spectroscopy. **Section V** is a brief summary.

## II. PHENOMENOLOGICAL DESCRIPTION

**A. Formulation of the General Problem**

The linear partial differential equation relating the mechanical displacement vector $U$ and elastic stress $\sigma_{ij}$ for a TMD has the form:

$$\rho \frac{\partial^2 U_i}{\partial t^2} + \frac{\partial \sigma_{ij}(x,t)}{\partial x_j} = \mu_{ij} \frac{\partial^2 P_j}{\partial t^2} + N_i[U_i], \qquad (1)$$

where $\rho$ is the density of material, $\mu_{ij}$ is the dynamic flexoelectric tensor [35], $P_j$ is the component of electric polarization, and $N_i[U_i]$ is the component of an elastic driving force causing the surface rippling. The physical origin of the force is the atomic reconstruction of the TMD surface, as confirmed by *ab initio* calculations [18 - 20, 31]. Within a continuum media approach, we used here, the force cannot be derived, but rather postulated. Mention that the applicability of a continuum media approach to ultra-thin TMD layers requires additional background provided by Berry et al. [41, 42] and later on by us [43].

The TMD can be mechanically free ("freestanding"), completely, or partially clamped to a rigid substrate. The displacement $U$ is zero at the clamped surfaces, e.g. at the bottom flat film-substrate interface, $x_3 = h$, shown in **Fig. 1c**, and the normal stress is absent at free surfaces, e.g. $\sigma_{ij} n_j |_S = 0$ at the rippled top surface in **Fig.**1c ($n_j$ is the j-component of the outer normal to the free surface S). Here $S$ is the SL free surface, and $h$ is the thickness of TMD in the form of ultra-thin film or flake.

Note that when the excitation pulse frequency ω is much smaller than the characteristic frequencies of TMD acoustic phonons, one can neglect time-dependent terms $\frac{\partial^2 U_i}{\partial t^2}$ and $\frac{\partial^2 P_j}{\partial t^2}$ in Eq.(1). Typically, the strong inequality $(\omega/\omega_a)^2 \ll 1$ is valid for scanning probe microscopy studies, where $\omega_a$ is the frequency of acoustic phonon, but is unlikely to hold for e.g. Raman spectroscopy.

Applying a continuum media approach, we further note that the system tends to the minimum of the Helmholtz free energy $F$ in the thermodynamic equilibrium, and $F$ consists of the bulk and surface parts, $F[P_i] = \int_V f_b[P_i]dV + \int_S f_S[P_i]ds$. The bulk and surface densities of the free energy, $f_b$ and $f_S$, are [44]:

$$f_b[P_i] = \begin{pmatrix} \frac{a_{ij}}{2} P_i P_j + \frac{b_{ijkl}}{4} P_i P_j P_k P_l + \frac{g_{ijkl}}{2} \frac{\partial P_i}{\partial x_j} \frac{\partial P_k}{\partial x_l} + q_{ijkl} P_i P_j u_{kl} \\ + \frac{f_{ijkl}}{2} \left( \frac{\partial P_i}{\partial x_j} u_{kl} - P_i \frac{\partial u_{kl}}{\partial x_j} \right) - P_i E_i + \frac{c_{ijkl}}{2} u_{ij} u_{kl} \end{pmatrix}, \quad (2a)$$

$$f_s[P_i] = \frac{a_{ij}^S}{2} P_i P_j - d_{ilk}^S P_i u_{lk}, \quad (2b)$$

Where $a_{ij}$, $b_{ijkl}$ are expansion coefficients on polarization powers, $c_{ijkl}$ is a tensor of elastic stiffness; $u_{kl} = \frac{1}{2} \left( \frac{\partial U_k}{\partial x_l} + \frac{\partial U_l}{\partial x_k} \right)$ is an elastic strain tensor; $d_{ilk}^S$ is a surface-induced piezoelectric tensor [21, 22], and $f_{ijkl}$ is a flexoelectric strain tensor. The components of $d_{ilk}^S$ normal to the TMD surface can be nonzero due to the absence of the out-of-plane mirror reflection operation [21, 22]. The electrostriction strain tensor $q_{ijkl}$ can be renormalized by the Maxwell stresses due to e.g. excess surface charges. The properties of the polarization gradient energy are described by a positively defined tensor $g_{ijkl}$. $E_i$ is the component of an electric field.

Since the spontaneous polarization is absent in most bulk TMD, the tensor $a_{ij}$ is positively defined, corresponding to lack of polar states in the bulk materials. However, its diagonal components can change the sign due to the renormalization ($a_{ij} \rightarrow a_{ij}^*$) by electrostriction and/or flexoelectric coupling, leading to the possibility of the second order phase transition at zero diagonal components, $a_{ii}^* = 0$. Since the coefficient $a_{ii}$ can change the sign, the stabilizing nonlinear terms $b_{ijkl} P_i P_j P_k P_l$ should be included in Eq.(2a), where the tensor $b_{ijkl}$ is positively defined. With these nonlinear terms, Eq.(2a) becomes similar to the Landau-Ginzburg-Devonshire (**LGD**) free energy used for the description of paraelectrics [37, 38, 40, 44].

For the case when the excitation frequency ω is much smaller than the optical one $\omega_{opt}$, i.e. for $\left(\omega/\omega_{opt}\right)^2 \ll 1$, the coupling between the stress $\sigma_{ij}$ and strain $u_{kl}$ is given by the generalized Hooke's law [45], obtained from the equation of state, $\sigma_{ij} = \frac{\partial f_b}{\partial u_{ij}}$,

$$\sigma_{ij} = c_{ijkl}\left(u_{kl} + f_{klmn}\frac{\partial P_m}{\partial x_n} + q_{klmn}P_m P_n\right). \quad (3a)$$

Allowing for the temporal Khalatnikov-type relaxation, minimization of Eq.(2a) yields the nonlinear dynamic equation for the polarization components, $P_i$,

$$\Gamma\frac{\partial P_i}{\partial t} + a_{ij}P_j + b_{ijkl}P_j P_k P_l - g_{ijkl}\frac{\partial^2 P_j}{\partial x_k \partial x_l} - q_{ijkl}P_l u_{jk} - f_{ijkl}\frac{\partial u_{jk}}{\partial x_l} - \mu_{ij}\frac{\partial^2 U_j}{\partial t^2} = E_i. \quad (3b)$$

The boundary conditions to Eq.(3) account for the flexoelectric effect [44, 46] and surface-induced piezoelectric coupling, $a_{ij}^S P_j - d_{ilk}^S u_{lk} + \left(g_{ijkl}\frac{\partial P_k}{\partial x_l} - F_{klij}\sigma_{kl}\right)n_j\bigg|_S = 0$, where **n** is the outer normal to the surface $S$, and $d_{ilm}^S$ is a surface-induced piezoelectric tensor [21, 22]. The strain is equal to the misfit strain $u_m$ at the clamped surface, e.g. $u_{11} = u_{22}|_S = u_m$. For a free surface $\sigma_{3i}|_S = 0$.

**B. Local Electric Fields**

The conjugate variable to the polarization behavior described by Eq.(1) is the electric field $E_i(\mathbf{r}, t)$, the product of which yields the energy term. Similar to classic ferroelectrics, the electric field is the sum of external field, $E_i^{ext}(\mathbf{r}, t)$, and internal depolarization field, $E_i^{dep}(\mathbf{r}, t)$, coming from the incomplete screening of the bound polarization charge by ambient free charges and/or (semi)conductive substrate. Depolarization often can include chemical effects, but here we ignore them assuming chemical stability of the materials.

The electric potential φ outside the TMD satisfies the homogeneous equation $\frac{1}{c^2}\frac{\partial^2 \phi}{\partial t^2} + \Delta\phi = 0$, and inhomogeneous equation inside it:

$$\frac{1}{c^2}\frac{\partial^2 \phi}{\partial t^2} + \varepsilon_f \Delta\phi = \frac{1}{\varepsilon_0}\frac{\partial P_i}{\partial x_i}, \quad (4)$$

where $\Delta = \frac{\partial^2}{\partial x_1^2} + \frac{\partial^2}{\partial x_2^2} + \frac{\partial^2}{\partial x_3^2}$ is a Laplacian, $\varepsilon_0$ is a universal dielectric constant, $\varepsilon_f$ is a relative permittivity of TMD film or flake.

Similarly to the analysis of elastic fields, the electric field can be found from quasi-static equations, if the excitation frequency ω is well below the optical range ($\omega \ll 10^{15} s^{-1}$), when the derivative $\frac{1}{c^2}\frac{\partial^2 \phi}{\partial t^2}$ can be neglected. For the case the field components $E_i$ are related to the electric potential $\phi$ in a conventional way, $E_i = -\partial\phi/\partial x_i$. Boundary conditions are the potential and electric displacement continuity at the TMD surface.

General solution of Eq. (4) can be represented as the sum of external (excitation) and internal (depolarization) fields. The electric potential in Fourier space, $\tilde{\phi}(\mathbf{k}, \omega)$, has the form:

$$\tilde{\phi}(\boldsymbol{k},\omega) = \tilde{\phi}_{ext}(\boldsymbol{k},\omega) + \frac{k_i \tilde{P}_i(\boldsymbol{k},\omega)}{\varepsilon_0 \varepsilon_f k^2} + \tilde{\phi}_c(\boldsymbol{k},\omega), \tag{5}$$

where $k^2 = \sqrt{k_1^2 + k_2^2 + k_3^2}$. The Fourier image of external field potential, $\tilde{\phi}_{ext}(\boldsymbol{k},\omega)$ is created by planar or probe-type electrodes, and so it has no poles, except for the first order pole at **k**=0, that reflects the long-range nature of quasi-static electric fields without any specific "local" scale.

The sum $\frac{k_i \tilde{P}_i(\boldsymbol{k},\omega)}{\varepsilon_0 \varepsilon_f k^2} + \tilde{\phi}_c(\boldsymbol{k},\omega)$ is a depolarization potential. The term $\frac{k_i \tilde{P}_i(\boldsymbol{k},\omega)}{\varepsilon_0 \varepsilon_f k^2}$ is a "bare" depolarization contribution, and $\tilde{\phi}_c(\boldsymbol{k},\omega)$ is the compensating potential required to satisfy the boundary conditions of electric potential and displacement continuity at the TMD interfaces. In the most general case, $\tilde{\phi}_c(\boldsymbol{k},\omega)$ is a linear integral-differential function of polarization. For the simplest case the shape of low dimensional TMD can be approximated by a film, or a very thin round plate or flake, the depolarization field can be described by a "local" depolarization factor [47], $\eta_{ij}(\boldsymbol{k},\omega)$, i.e.

$$\tilde{E}_i(\boldsymbol{k},\omega) \approx \tilde{E}_i^{ext}(\boldsymbol{k},\omega) - \frac{\eta_{ij}(\boldsymbol{k},\omega)}{\varepsilon_0 \varepsilon_b} \tilde{P}_j(\boldsymbol{k},\omega). \tag{6}$$

We note that high depolarization factors effectively preclude the formation of a ferroelectric state in nanosized ferroics [44]. The small depolarization factors, used in further calculations, correspond to a high degree of polarization screening by the proper electronic conductivity of TMD, or ambient free charges and/or conductive substrate.

The formulation of the coupled problem (1)-(6) does not allow for solution in general case of an arbitrary low-dimensional TMD. This is similar to classical ferroelectric materials, in which depolarization field renders the electrostatic problem non-local, drives the emergence of vortex states, and generally define the unique physics of these systems. Similarly to ferroelectric materials, to gain insight into the polar physics of 2D TMDs, we consider several special cases.

### C. Frequency Limits of Polarization Dynamics

The quasi-stationary solution of Eq.(3) for the polarization component $P_i(\boldsymbol{r},t)$ can be found in an adiabatic approximation, applicable when the corresponding Landau-Khalatnikov time $\tau_K = \Gamma/|a_{ii}|$ is much smaller than the characteristic times of external electric field changes, i.e. the strong inequality $\tau_K \omega \ll 1$ is fulfilled. Only in this case one can regard that the polarization changes immediately follow the applied voltage pulses, however the estimates of $\tau_K$ give very broad range for different ferroics.

For a proper ferroelectric far from the ferroelectric Curie point (e.g. at room temperature), $|a_{ii}| \cong (10^7 - 10^8)$ Jm/C$^2$, so the Landau-Khalatnikov time can be estimated as $\tau_K \cong (10^{-6} - 10^{-9})$s, and the low-frequency limit means that $\omega \ll (10^6 - 10^9)$1/s. The estimates of $\tau_K$ for quantum paraelectrics with high dielectric susceptibility and especially for relaxors with continuous spectra of relaxation times, can lead to much higher values, $\tau_K > (10^{-6} - 10^{-3})$s and so it must be $\omega \ll (10^3 - 10^6)$1/s.

While these estimates are universally derived for the three-dimensional systems, the segregation of dopants is likely in low-dimensional systems. Furthermore, there can be large scale ripples. Finally, extrinsic adsorbed dopants can effectively contribute to the relaxation time spectra. Therefore, we assume that there are a lot of possible intrinsic and extrinsic slowing down mechanisms, and consider a general case, for which both inequalities $\tau_K \omega \ll 1$ or $\tau_K \omega \gg 1$ can be valid a priory for the considered SL-TMDs.

## 3. THE PROPERTIES OF FLEXO-INDUCED POLARIZATION

### A. Linear Response of the Strain-Induced Polarization

As the first step analysis we regard that the excitation field is much smaller than the coercive one, and look for analytical solutions of Eq.(3b) in the simplest one-component and one-dimensional approximation. Allowing for Eq.(6) and omitting all tensorial notations, Eq.(3b) reads:

$$\Gamma \frac{\partial P}{\partial t} + \left[a + \frac{\eta}{\varepsilon_0 \varepsilon_b} + qu(x)\right] P + bP^3 - g \frac{\partial^2 P}{\partial x^2} = E - f \frac{\partial u}{\partial x} + \mu \frac{\partial^2 U}{\partial t^2}, \quad (7a)$$

where the gradient coefficient is introduced as $g = g_{44} - \frac{f^2}{c}$, as renormalized by the flexoelectric coupling. We consider the case $g > 0$, that is below the upper limit for the static flexoelectric effect, which is considered in Refs.[48, 49]. Since a bulk TMD is not a ferroelectric, the coefficient $a > 0$. The depolarization factor $\eta > 0$, because the depolarization effects can only impede the emergence of a ferroelectric phase. The condition for the nonlinear coefficient $b > 0$ is necessitated by the stability of material with free energy Eq.(2a). The signs of the electrostiction, static and dynamic flexoelectric coefficients, $q$, $f$ and $\mu$, can be arbitraty. External electric field $E$ is taken periodic and homogeneous, $E(\omega) = E_0(\omega) \sin(\omega t)$.

To analyze the polarization responses in the material, we assume that a small displacement $U(x)$ and strain $u(x) = \frac{\partial U}{\partial x}$, caused by the surface rippling, has the form of a static harmonic function,

$$U(x) = U_0(\omega) \cos(k_s x + \psi), \qquad u(x) = u_m + u_0(\omega) \sin(k_s x + \psi), \quad (7b)$$

where $k_s$ is the modulation wavenumber, $u_0 = -k_s U_0$ is the strain amplitude, and $\psi$ is an arbitrary constant phase. Note that the rippling is postulated to be induced by the elastic force $N_i[U_i]$ originating from the surface reconstruction [18 - 20, 31] or interaction with external or internal periodic instabilities [50].

Since $|k_s| > 0$, and given that we are interested in a periodic linearized solution with the wavenumbers $k = \pm k_s$, the linear polarization response to a small electric excitation and strain can be roughly estimated in $\{x, \omega\}$-space as:

$$P(x, \omega) \approx P_S + \frac{E(\omega) - (fk_s^2 + \mu \omega^2) U_0(\omega) \cos(k_s x + \psi)}{i\omega\Gamma + a + \frac{\eta}{\varepsilon_0 \varepsilon_b} + qu_m + gk_s^2 + 3bP_S^2}. \quad (8)$$

The derivation of Eq.(8) is given in **Appendix A**.

The spontaneous polarization is absent for bulk TMDs, however the irreversible polarization $P_S$ may appear in a SL-TMD due to the out-of-plane surface piezoelectric effect coupled with the misfit-induced (possibly quasi-homogeneous) local strain $u_m$ as $P_S \sim \frac{d_{13}^S}{a_{33}^S} u_m$. Naturally, the misfit strain is absent ($u_m = 0$) for a freestanding TMD film/flake or for a "matched" substrate with an effective lattice constant coinciding with TMD, and so the first term in Eq.(8) is absent in the case.

As an illustration, the linear one-component response, i.e. the out-of-plane polarization Eq.(8) appeared in response to the small strain Eq.(7b), is shown in **Fig. 2**. The linear responses are calculated for negative, zero and positive flexoelectric coefficient $f$, different frequencies $\omega$ of external excitation, negative, zero and positive misfit strains $u_m$. All physical values and parameters were dimensioned to characteristic values, and $x$-scale is $R_S = \frac{1}{k_S}$. The small dimensionless depolarization factors $\frac{\eta}{a\varepsilon_0\varepsilon_b} \ll 1$, used in the calculations, correspond to a high degree of polarization screening by the intrinsic conductivity and/or conductive substrate. The amplitude $u_0$ depends on the elastic boundary conditions, it is taken significantly bigger for a freestanding SL-TMD in comparison with the layer clamped to a rigid substrate (compare the amplitude of dashed curves in **Fig. 2a** and **Fig. 2c**).

**Fig. 2a** shows that the flexoelectric coupling induces the polarization response to the oscillating local strain in a freestanding SL-TMD, that is phase shifted on $-\frac{\pi}{2}$ (see blue curves for $f < 0$) or $+\frac{\pi}{2}$ (see red curves for $f > 0$) with respect to the local strain phase. **Fig. 2b** shows that the polarization response rapidly decreases when the excitation frequency exceeds the characteristic value $\omega\tau_K = 1$.

**Fig. 2c** indicates that the dependence on misfit strain is observed for a wide range of frequencies and $u_m$ values, at that we take enough high value of the dimensionless surface-induced piezoelectric coefficient ($d^S = 1$). Actually, blue and red curves calculated for high compressive (-5%) and tensile (+5%) strains are significantly shifted upward, as anticipated for $d^S > 0$. The black curve in **Fig. 2c** is calculated for a zero-misfit strain, corresponding to the so-called "matched substrate". Solid curves in **Fig. 2d**, calculated for a tensile misfit strain +5%, shows that the polarization response of a clamped SL-TMD relatively slowly decreases with the excitation frequency increase, even when it exceeds the characteristic value $\omega\tau_K = 1$. Dashed curves in **Fig. 2d**, calculated for a compressive misfit strain -5%, shows that the polarization response of a clamped SL-TMD starts to significantly decrease with the excitation frequency increase, when it exceeds the characteristic value $\omega\tau_K = 1$.

The results presented in **Fig. 2** demonstrate that the flexoelectric effect determines the amplitude of the linear polarization response, and shifts its phase on 90 degree. Thus, we can conclude that the flexoelectric effect, that is widespread in contrast to the surface-induced piezoelectric effect, can be responsible for the origin polarization response of the SL-TMD if the local strain is caused by the surface rippling, corrugation or bending.

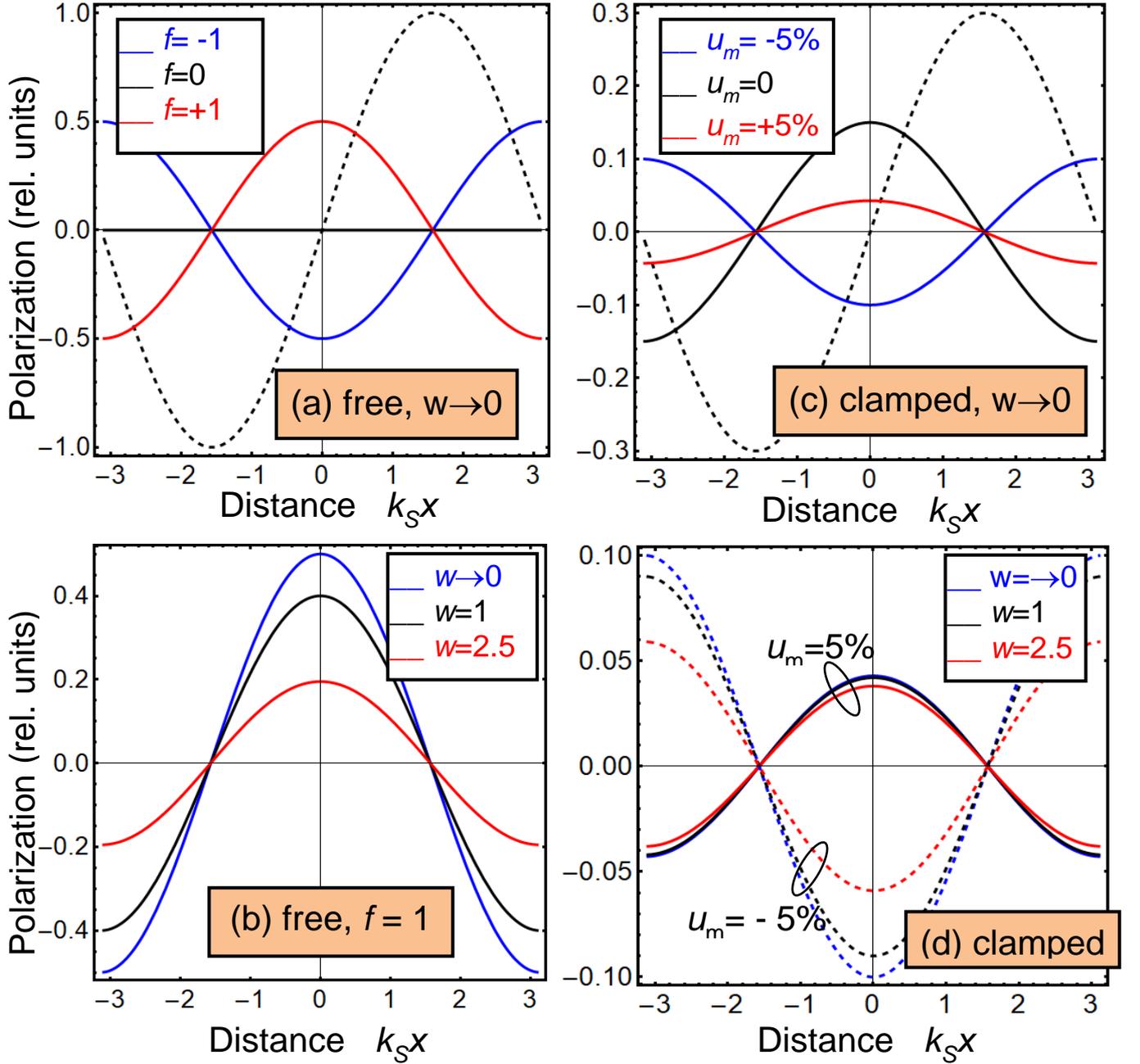

**FIGURE 2. Linear polarization response.** Polarization distribution $P(x)$ (solid curves) induced by the local strain $u(x)$ in the form of static harmonic ripples (dashed curve). The polarization curves of a freestanding SL-TMD is shown in plots **(a, b)**, and the curves in plots **(c, d)** correspond to the SL-TMD clamped to a rigid substrate. Solid curves in the plots **(a)** are calculated for several values of the dimensionless flexoelectric coefficients $f = -1$ (blue curves), 0 (black curves) and +1 (red curves) and $\omega \tau_K = 0.001$. Solid curves in the plot **(b)** are calculated for several dimensionless frequencies $\omega \tau_K = 0.001$ (blue curves), 1 (black curves) and 2.5 (red curves), and $f = 1$. Solid curves in the plot **(c)** are calculated for several misfit strains $u_m = -5\%$ (blue curves), 0 (black curves) and $+5\%$ (red curves), $f = 1$ and $\omega \tau_K = 0.001$. Curves in the plot **(d)** are calculated for the dimensionless frequencies $w = \omega \tau_K$, which are equal to 0.001 (blue curves), 1 (black curves) and 2.5 (red curves), $u_m = +5\%$ (solid curves) and $u_m = -5\%$ (dashed

curves), $f = 1$. External field is enough small. Other dimensionless parameters are normalized, $\frac{gk_S^2}{a} = 1$, $\frac{q}{a} = 100$, $\frac{d_{13}^S}{a_{33}^S} = 1$, $\frac{b}{a} = 0.1$, $\frac{\eta}{a\varepsilon_0\varepsilon_b} = 10^{-2}$ and $\mu = 0$.

Without external field ($E = 0$), the variation of the polarization in a full tensorial case is proportional to the convolution of the local strain tensor with the surface-induced piezoelectric and flexoelectric tensors, $\widetilde{\delta P}_i(\mathbf{k},\omega) \sim ik_l f_{ijkl} \tilde{u}_{jk}(\mathbf{k},\omega)$. Since $|f_{ijkl}| < 10^{-10}$m$^3$/C [3 - 32, 51] and the strain of SL-TMD unlikely exceed 5% (maximal strain reported in SL-TMDs can be 11%) the linear polar response is detectable, while rather small, except for the possible poles of the dielectric susceptibility in **k**-space.

### B. Linear Dielectric Susceptibility

The $\{k,\omega\}$-spectrum of linear dielectric susceptibility $\chi(k,\omega)$ is renormalized by the local depolarization effects, misfit strain and "acting" component of the rippling as

$$\chi(k,\omega) \approx \frac{a + \frac{\eta(k,\omega)}{\varepsilon_0\varepsilon_b} + qu_m - qu_0\left(1 - \frac{k_S^2}{2(k^2+k_S^2)}\right) + gk^2 + 3bP_S^2 - i\omega\Gamma}{\left(a + \frac{\eta(k,\omega)}{\varepsilon_0\varepsilon_b} + qu_m - qu_0\left(1 - \frac{k_S^2}{2(k^2+k_S^2)}\right) + gk^2 + 3bP_S^2\right)^2 + (\omega\Gamma)^2}. \tag{9}$$

Derivation of Eq.(9) is given in **Appendix A**. The expression for the dielectric response in $\{x,t\}$-space can be obtained from Eq.(9) using inverse $\{k,\omega\}$-Fourier transforms, and, in general case, the procedure can give a generalized response function, that can be negative in some x-points, similarly to an "effective" negative permittivity [52].

The condition in $\{k,\omega\}$-space

$$\text{Re}[\chi(k,\omega)] < 0, \tag{10}$$

indicates the possibility of the strain-induced transition to a ferroelectric-like phase with a robust bistable out-of-plane polarization. Since an external electric field has no poles except for the pole at *k*=0, the scale of local dielectric response is defined by the minima or zeros of inverse susceptibility, $\frac{1}{\chi(k,\omega)} = 0$.

The real and imaginary parts of the dielectric susceptibility spectra, Re[$\chi(k,\omega)$] and Im[$\chi(k,\omega)$], are shown in **Fig. 3** for a freestanding and clamped SL-TMD, for several values of frequencies and misfit strains. The *k*-scale is shown in the units of the bulk' correlation length $R_c = \sqrt{g/a}$, that is chosen equal to the rippling period $R_S$, because we obtained that the equality $R_S = R_c$ leads to the maximal correlation between the periodic strain and polarization. The susceptibility spectrum is an even function of *k*, as anticipated from the approximate Eq.(9).

For a freestanding SL-TMD two positive and two negative sharp peaks of Re[$\chi(k,\omega)$] exist in the *k*-range $-3 < kR_c < 3$ for dimensionless frequency range, $\omega\tau_K = (0.001 - 0.1)$, which is relatively low in comparison with a phonon relaxation time $\tau_K \approx \frac{\Gamma}{|a|}$. [see **Fig. 3a**]. It is seen that the region of Re[$\chi(k,\omega)$] double peaks corresponds to the sharp negative peaks of Im[$\chi(k,\omega)$] located in the same $\{k,\omega\}$ range [see **Fig. 3b**]. For $\omega\tau_K = 1$ the peaks of dielectric susceptibility spectrum smear significantly.

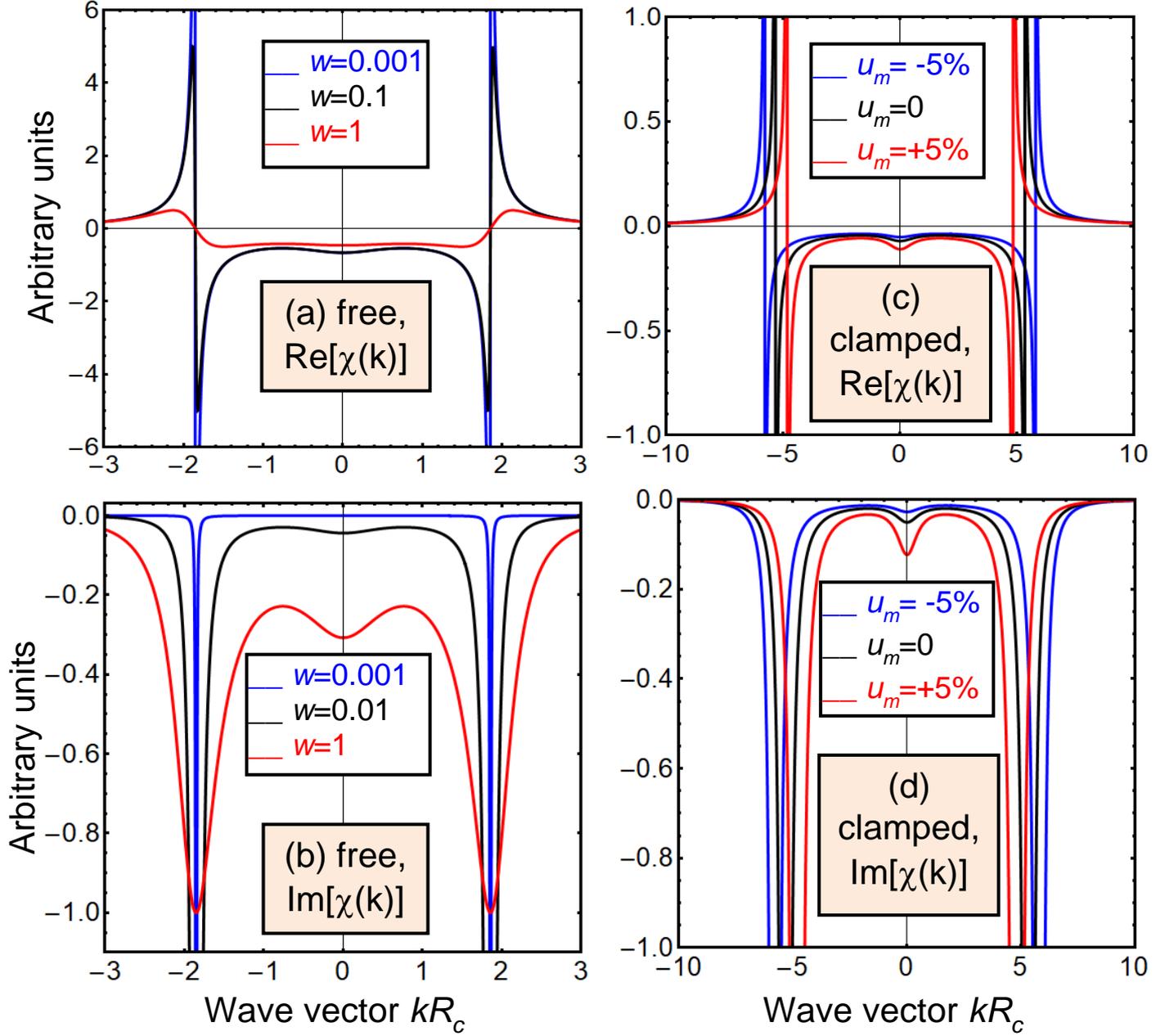

**FIGURE 3. Linear dielectric susceptibility spectrum.** The real (**a, c**) and imaginary (**b, d**) parts of the linear dielectric susceptibility spectrum, $\chi(k,\omega)$, calculated for a freestanding SL-TMD (**a, b**), and for a SL-TMD clamped to a rigid substrate (**c, d**). Curves in the plots (**a, b**) are calculated for different dimensionless frequencies $w = \omega\tau_K$, which are equal to 0.001 (blue curves), 0.1 (black curves) and 1 (red curves). Curves in the plots (**c, d**) are calculated for different misfit strains $u_m = -5\%$ (blue curves), 0 (black curves) and $+5\%$ (red curves), and frequency $w = 0.1$. Parameters $f = 1$, $d^S = 1$, and $R_S = R_c$. Other parameters are the same as in **Fig. 2**.

**Figures 3c-d** illustrate that misfit strains shift the maxima of real and imaginary parts of the dielectric susceptibility spectrum within the k-range $-10 < kR_c < 10$ for a SL-TMD clamped to a rigid substrate.

Black curves in the figure are calculated for a zero-misfit strain, corresponding to the so-called "matched substrate". It appears that a tensile strain leads to the left *k*-shift of the $\text{Re}[\chi(k,\omega)]$ and $\text{Im}[\chi(k,\omega)]$ peaks, and the shift is significant not only at l ow frequencies $\omega \tau_K \leq 0.1$ (see red curves calculated for high tensile strains +5%), but also for a much wider range of material parameters (not shown). A compressive strain leads to a significant right *k*-shift of the $\text{Re}[\chi(k,\omega)]$ and $\text{Im}[\chi(k,\omega)]$ peaks (see blue curves calculated for high compressive strains -5%). Due to the clamping effect the peaks of $\text{Re}[\chi(k,\omega)]$ in a SL-TMD at substrate are much sharper and higher at $\omega \tau_K \geq 0.1$ than the same peaks in a freestanding SL-TMD (compare black curves in **Fig. 3c-d** and **3a-b**). At the same time the imaginary part, $\text{Im}[\chi(k,\omega)]$, is multiplied by a factor 100.

Notably that the linear dielectric susceptibility spectrum is tightly related with the correlation function of polarization fluctuations [53, 54]. Corresponding anisotropic correlation lengths $R_{ij}^C$ can be roughly estimated as:

$$\frac{1}{R_{ii}^C(\mathbf{k},\omega)} \cong \frac{1}{R_C}\sqrt{\frac{g_{11}}{g_{ii}^{eff}|a_{11}|}}\left|a_{ii} + Q_{iikl}c_{klmn}u_{mn}^s(\mathbf{k},\omega) + \frac{\eta_{ii}(k,\omega)}{\varepsilon_0\varepsilon_b} + b_{iilm}P_m^s P_l^s\right|, \quad (11)$$

where the summation over "*i*" is absent in Eq.(11), and $g_{ii}^{eff}$ is a combination of gradient coefficients $g_{iilm}^{eff}$, renormalized by a flexoelectric coupling [49], and the "bulk" correlation length $R_c$ is introduced as $R_c = \sqrt{\frac{g_{11}}{|a_{11}|}}$. Approximate numerical estimates of the correlation lengths depend strongly on temperature, local strains and depolarization factors. It ranges from nm to hundreds of nm, and apparently diverges when the system approaches the second order phase transition to a ferroelectric phase. Expression (11) indicates on a well-known fact, that the points of maximal linear dynamic susceptibility in **r**-space are the most probable candidates for the regions of maximal local response. The first candidates for these points are the immediate vicinity (about several $R_{ii}^C$) of the curved edges of the film, flake, corrugations or the structural antiphase boundaries, or elastic twin walls with giant linear susceptibility.

### C. Emergence of a Bistable Polarization States in a Non-Uniform Electric Field

We further explore the emergence of the spontaneous bending-induced polarization and especially its bistable states in low-dimensional TMDs. The interest is related with the fact, that the flexo-induced changes can be easier detected experimentally with the strain increase. Actually, to "observe" the local strain-induced polarization in low-dimensional TMDs, piezoresponse force microscopy (**PFM**) operating within frequency range ($10^3 - 10^6$) Hz can be used [55, 56, 57]. The spectroscopic local probing detects the surface displacement and forces at the nanoscale, and the dynamics of local electromechanical response is highly sensitive to the local fields [58].

Since the electric fields and elastic strains induced by a PFM probe can be very high, the approximate expressions (7)-(9), corresponding to a linear local response, are inapplicable a priory for comparison of the proposed theoretical model with e.g. PFM experiments. One should take into account the cubic (or higher) nonlinearity and, not less important, the nonlocality of the studied polar response.

In particular, within the framework of one-dimensional approximation, one should solve two coupled nonlinear partial differential equations for polarization and displacement components, $P(x)$ and $U(x)$, obtained from Eqs.(1), (3) and (7):

$$\frac{\partial^2 U}{\partial x^2} = \frac{1}{c} N[U] - f \frac{\partial^2 P}{\partial x^2} - 2qP \frac{\partial P}{\partial x}, \quad (12a)$$

$$\left[a + \frac{\eta}{\varepsilon_0 \varepsilon_b} + q \frac{\partial U}{\partial x}\right] P + bP^3 - g \frac{\partial^2 P}{\partial x^2} = E - f \frac{\partial^2 U}{\partial x^2}, \quad (12b)$$

supplemented with periodic (or antiperiodic) boundary conditions,

$$U(L) = U(\psi - L), \quad \frac{\partial U}{\partial x}(L) = \frac{\partial U}{\partial x}(\psi - L), \quad P(L) = P(\psi - L), \quad \frac{\partial P}{\partial x}(L) = \frac{\partial P}{\partial x}(\psi - L), \quad (12c)$$

imposed at the ends of the computation cell of width $2L$; $\psi$ is a constant phase. To model the appearance of periodic ripples at the reconstructed free surface we use the simple function of the force,

$$\frac{1}{c} N[U] = -\gamma (U - U_0) + \epsilon (U - U_0)^3, \quad (12d)$$

where the elastic modulus $c$ is positive, and strengths constants are not negative, $\gamma \geq 0$ and $\epsilon \geq 0$. The force consists of a linear contribution, $-\gamma(U - U_0)$, that is a usual Hooke-type elastic force from the equilibrium position $U = U_0$, and the third power nonlinearity expansion of nonlinearity, $\epsilon(U - U_0)^3$.

Finite element modeling (**FEM**) results, which are presented below, assure us that two different cases, namely the "weak" and "strong" electromechanical coupling, should be considered separately.

The **"weak" coupling** case is realized if $a + \frac{\eta}{\varepsilon_0 \varepsilon_b} + qu_0 > 0$. In this case the bending-induced and surface-induced polarization is small, and so the nonlinear term $bP^3$ does not influence significantly the solution of Eq.(12b), and simultaneously $P$-dependent terms can be neglected in Eq.(12a) as the first approximation. The trial functions for the displacement and strain, which well fit the numerical solution, have the form of elliptic sinus and its $x$-derivative:

$$U(x) = U_0 + U_S \sqrt{\frac{2n}{1+n}} \, \text{sn}\left(\frac{xk_s}{\sqrt{1+n}}, n\right), \quad (13a)$$

$$u(x) = \frac{\partial U}{\partial x} \approx U_S k_s \frac{\sqrt{2n}}{1+n} \, \text{cn}\left(\frac{xk_s}{\sqrt{1+n}}, n\right) \text{dn}\left(\frac{xk_s}{\sqrt{1+n}}, n\right). \quad (13b)$$

where $\epsilon > 0$, the amplitude $U_S = \sqrt{\gamma/\epsilon}$, parameter $0 < n < 1$, and $k_s = \sqrt{\gamma}$. At $\epsilon = 0$ the elliptic sinus transforms to a harmonic function, and so $u(x) = u_m + u_0 \sin(k_s x)$, where $u_0$ is an arbitrary amplitude. If the term $qu(x)$ becomes compatible with $a + \frac{\eta}{\varepsilon_0 \varepsilon_b}$ in Eq.(12b), but still smaller, the approximate linearized solution (8) of this equation is a suitable trial function for a FEM [see details in **Appendix A**]:

$$P(x) \approx \frac{1}{gk_s^2 + a + \frac{\eta}{\varepsilon_0 \varepsilon_b} + qu_m} \left[E - f \frac{\partial u}{\partial x}\right]. \quad (13c)$$

The weak coupling regime is non-hysteretic with respect to external electric field $E$, since the polarization (13c) is linearly proportional to the external electric field and strain derivative. The case of weak coupling is shown in **Fig. 4a** and **Fig. 4b.**

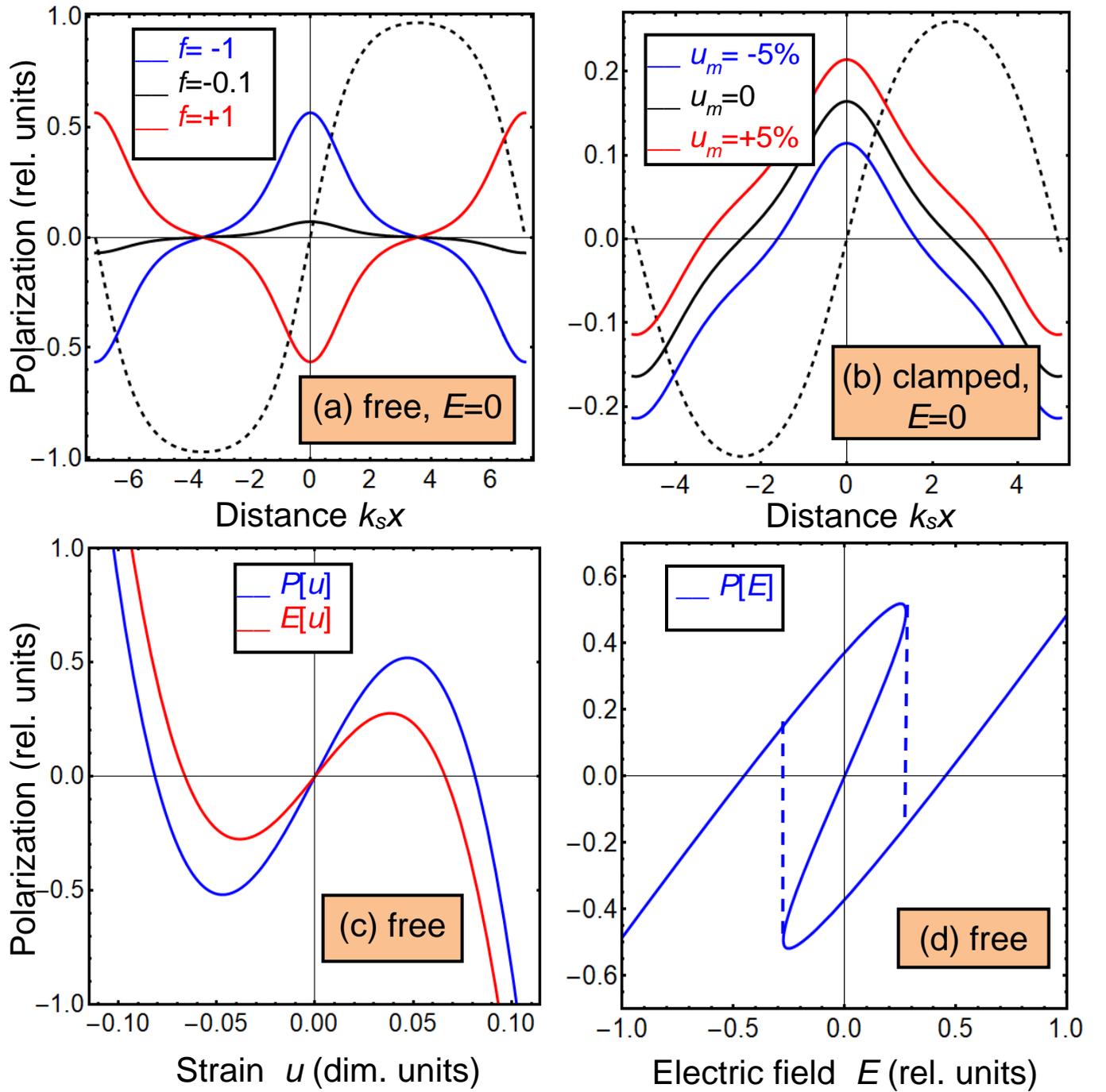

**FIGURE 4. Static polarization response of a SL-TMD.** Polarization distribution $P(x)$ (solid curves) induced by the local strain $u(x)$ in the form of static un-harmonic ripples (dashed curve). The SL-TMD is freestanding **(a)** or clamped to a rigid substrate **(b)**. Solid curves in the plots **(a)** are calculated for three values of the dimensionless flexoelectric coefficients $f = -1$ (blue curves), 0 (black curves) and +1 (red curves) and very low dimensionless frequency $\omega \tau_K = 0.001$. Solid curves in the plot **(b)** are calculated for three misfit strains $u_m = -5\%$ (blue curves), 0 (black curves) and +5% (red curves), $f = 1$ and $\omega = 0$. Parameters $\gamma = 1$, $\epsilon = 100$, $f = 1$ and $d = 10R_c$. Other dimensionless parameters are the same as in **Fig. 3**. External electric field is absent. Average polarization of the freestanding SL in dependence on strain **(c)** and external electric field **(d)**.

**Figures 4a-b** illustrate the main features of the freestanding and clamped SL-TMD nonlinear static polarization response. Shown is the appearance of the quasi-static out-of-plane polarization response to a periodic strain with high enough amplitude and inharmonic profile. The amplitude $u_0$ depends significantly on the elastic boundary conditions in a harmonic regime: it appeared significantly bigger for a freestanding SL-TMD in comparison to the layer clamped to a rigid substrate (compare the amplitude of dashed curves in **Fig. 4a** and **Fig. 4b**). However, the spontaneous polarization, induced by a misfit strain and symmetry-sensitive surface-induced piezoelectric coupling, is electret-like and irreversible [21, 22].

The polarization profiles are strongly nonlinear (see solid curves in in **Figs. 4a-b**), and the nonlinearity is caused by the anharmonicity of strain response at $E=0$ (see dashed curves in in **Figs. 4a-b**). It appeared that the ferroelastic-type hysteresis is possible is this regime, which becomes the most pronounced in the case of a freestanding SL-TMD. The physical origin of the hysteresis effect can be explained using a toy model with an electrostriction force produced by a non-uniform electric field. Such fields can be induced by a charged PFM probe, or/and by local disorder (charged defects). The coupled system (12), linearized with respect to polarization, acquires the form:

$$-\gamma U + \epsilon U^3 - \frac{\partial^2 U}{\partial x^2} = f \frac{\partial^2 P}{\partial x^2}, \qquad P - R_c^2 \frac{\partial^2 P}{\partial x^2} \approx \chi\left(E - f \frac{\partial^2 U}{\partial x^2}\right), \tag{14}$$

where $\chi = \frac{1}{a + \frac{\eta}{\varepsilon_0 \varepsilon_b}}$ is an "effective" linear dielectric susceptibility, and we introduce the substitution $(U - U_0) \to U$ for the sake of simplicity. Using decoupling approximation for a small polarization, we omit the terms proportional to a small flexoelectric coefficient in the right-hand-side of the first of Eq.(14), i.e. we set that $\frac{\partial^2 P}{\partial x^2} \approx \chi\left(1 + \frac{R_c^2}{d^2}\right)\frac{\partial^2 E}{\partial x^2}$, but remained the term $f\frac{\partial^2 U}{\partial x^2}$ in the equation for polarization. Next, assuming that one can use the following expressions for the non-uniform electric field and its second derivative at the TMD surface, $E(x) = \frac{E_0 x}{\sqrt{x^2 + d^2}}$ and $\frac{\partial^2 E}{\partial x^2} \approx \frac{-E_0 x}{\sqrt{(x^2+d^2)^3}} \approx \frac{E(x)}{d^2}$ at $x^2 \ll d^2$, where $d$ is either the effective size of PFM tip, or the size of charged defect [37, 55 - 57], one can further simplify Eqs.(14) and rewrite them via the displacement, as a parametric variable [see **Appendix A**], namely,

$$E[U] \approx -\frac{d^4}{f\chi(d^2 + R_c^2)}\left(-\gamma U + \epsilon U^3 - \frac{\partial^2 U}{\partial x^2}\right), \tag{15a}$$

$$P[U] \approx -\frac{d^2}{f}\left(-\gamma U + \epsilon U^3 - \frac{\partial^2 U}{\partial x^2}\right) - \chi f \frac{\partial^2 U}{\partial x^2}. \tag{15b}$$

The static dependences of the field $E[u]$ and polarization $P[u]$ vs. the strain $u \approx \frac{U}{h}$ are shown in **Fig. 4c** for a freestanding SL-TMD. It is seen from the figure the dependences have the form of S-curves, which extrema (minima and maxima) are shifted with respect to one another. The positions of the extrema are different for $E[u]$ and $P[u]$, and shifted from the points $u_S = \pm\sqrt{\gamma/\epsilon h^2}$ due to the presence of the elastic gradient terms, $\frac{\partial^2 U}{\partial x^2}$ and $\chi f \frac{\partial^2 U}{\partial x^2}$, in Eqs.(15). The terms $\sim \frac{\partial^2 U}{\partial x^2}$, which are present in both equations (15), shift the extrema in the same way closer to the axis $u = 0$. However, the term proportional to $\chi f \frac{\partial^2 U}{\partial x^2}$, standing alone in equation (15b) for $P[u]$ and being absent in Eq.(15a) for $E[u]$, additionally shifts the P-extrema position from $u_S$, approximately $u_S \to \pm\sqrt{(\gamma - \chi f^2 k_s^2 d^{-2})/\epsilon h^2}$. Since the S-curves of $E[u]$ and $P[u]$ have different extrema positions, the apparent hysteresis-like dependence can originate at the parametric curve $P[E]$, and we are interested exactly in the case. Actually, the pronounced S-shape hysteresis curve for $P[E]$, shown in **Fig. 4d**, can be interpreted as the origin of bistable and reversible by an electric field ferroelectric-type polarization in a freestanding SL-TMD with a rippled surface. Since the term, $\chi f \frac{\partial^2 U}{\partial x^2}$, responsible for the P-hysteresis, is proportional to the flexoelectric coefficient $f$, we deal with the flexoelectric origin of the bending-induced ferroelectricity. To background this semi-analytical result detailed FEM simulations for entire range of material parameters inherent to SL-TMDs and its comparison to experiment are required. We should underline, that within the scope of used approximations, the bistable polarization states appear in a non-uniform electric field only.

The **"strong" coupling** case can be realized if $a + \frac{\eta}{\varepsilon_0 \varepsilon_b} + qu_0 < 0$. The case corresponds to a true ferroelectric coupling in the system. However an estimates of material parameters in Eqs.(12) rise some doubts in the realization of strong coupling regime for e.g. SL-MoS$_2$ or SL-WS$_2$, while the possibility is not excluded in general. The case will be considered elsewhere.

## 4. CONDITIONS OF THE LOCAL RESPONSE ENHANCEMENT IN RAMAN SPECTROSCOPY AND PLASMONICS

The surfaced-enhanced vibrational spectroscopy methods, such as surface-enhanced (**SE**) Raman scattering (**RS**) probing by laser excitation in a visible range and surface enhanced infra-red abortion (**SEIRA**), which obtain unique information about the IR region or vibrational modes in the IR-frequencies, can be used to visualize the polar phenomena in graphene and graphene-like low-dimensional TMDs [59, 60, 61, 62]. To reach a high resolution and signal detection level using the surfaced-enhanced vibrational spectroscopy, the real part of the dielectric susceptibility spectra should be negative and as high as possible. The explanation of the fact is that the scattering methods are extremely sensitive to the local electric fields. In turn, the local fields can lead to the giant enhancement of observed response under the specific plasmon "resonant" (or even non-resonant with essential density of states of charge carriers) or phonon resonant conditions. For instance, for SERS or SEIRA the enhancement condition is expressed via the gain factor [63]:

$$g(k,\omega) = -\frac{\text{Re}[\chi(k,\omega)]}{|\text{Im}[\chi(k,\omega)]|} \gg 1. \tag{16}$$

The candidates for high gain factors in Eq.(16) correspond to zeros and special points (poles or divergence at $\omega\Gamma \to 0$) of the linear dynamic susceptibility **k**-spectra (9). From Eq.(9), the susceptibility is maximal (if $\omega\Gamma > 0$) or diverges (if $\omega\Gamma \to 0$) when $a + \frac{\eta(k,\omega)}{\varepsilon_0 \varepsilon_b} + qu_m - qu_0\left(1 - \frac{k_s^2}{2(k^2+k_s^2)}\right) + gk^2 + 3bP_s^2 \to 0$.

The spectra of the gain factor, $g(k,\omega)$, calculated for a freestanding and clamped SL-TMD at several frequencies and misfit strains, are shown in **Fig. 5a** and **5b**, respectively. The *k*-scale and other parameters are the same as in **Fig. 3**.

Actually, the region of high gain factors, $g(\boldsymbol{k},\omega) \gg 1$, responsible for the amplification of the signals in the IR, SERS and SEIRA methods [59 - 62], exists and reaches high values ~(5 – 2000) for the freestanding SL-TMD [see blue and black curves in **Fig. 5a**]. As it was expected, the condition $g(k,\omega) \gg 1$ can be reached easier at low frequencies within the range $\omega\tau_K = (0.001 - 0.1)$; and the gain factor decreases rapidly for $\omega\tau_K > 0.1$.

**Figs.5b** illustrates how the misfit strain can change the gain factor spectrum for a SL-TMD at different substrates. High compressive strains increase the gain factor (blue curves for $u_m = -5\%$), and tensile strains decreases it (red curves for $u_m = +5\%$) in comparison with the gain factor calculated for the case of matched substrate (black curves for $u_m = +5\%$). The gain peaks of a clamped SL-TMD are significantly wider and higher in comparison with the gain peaks of a freestanding SL for every of three frequencies (tree groups of curves marked by an ellipse in **Fig. 5b** with a corresponding single curve in **Fig. 5a**). Appeared, that the trend is typical not for the dimensionless parameters used in the figure, but also for the wider range of material parameters. Hence, we can conclude that the partial mechanical clamping of a low-dimensional TMD to a substrate can essentially increase the gain factor, and the result can be readily verified experimentally. As expected, the gain factor be principally different for a dielectric and conductive substrate, because the susceptibility depends on the dimensionless depolarization factor, $\frac{\eta}{a\varepsilon_0\varepsilon_b}$, which value is defined by a degree of polarization screening by internal and ambient charges, and/or substrate conductivity.

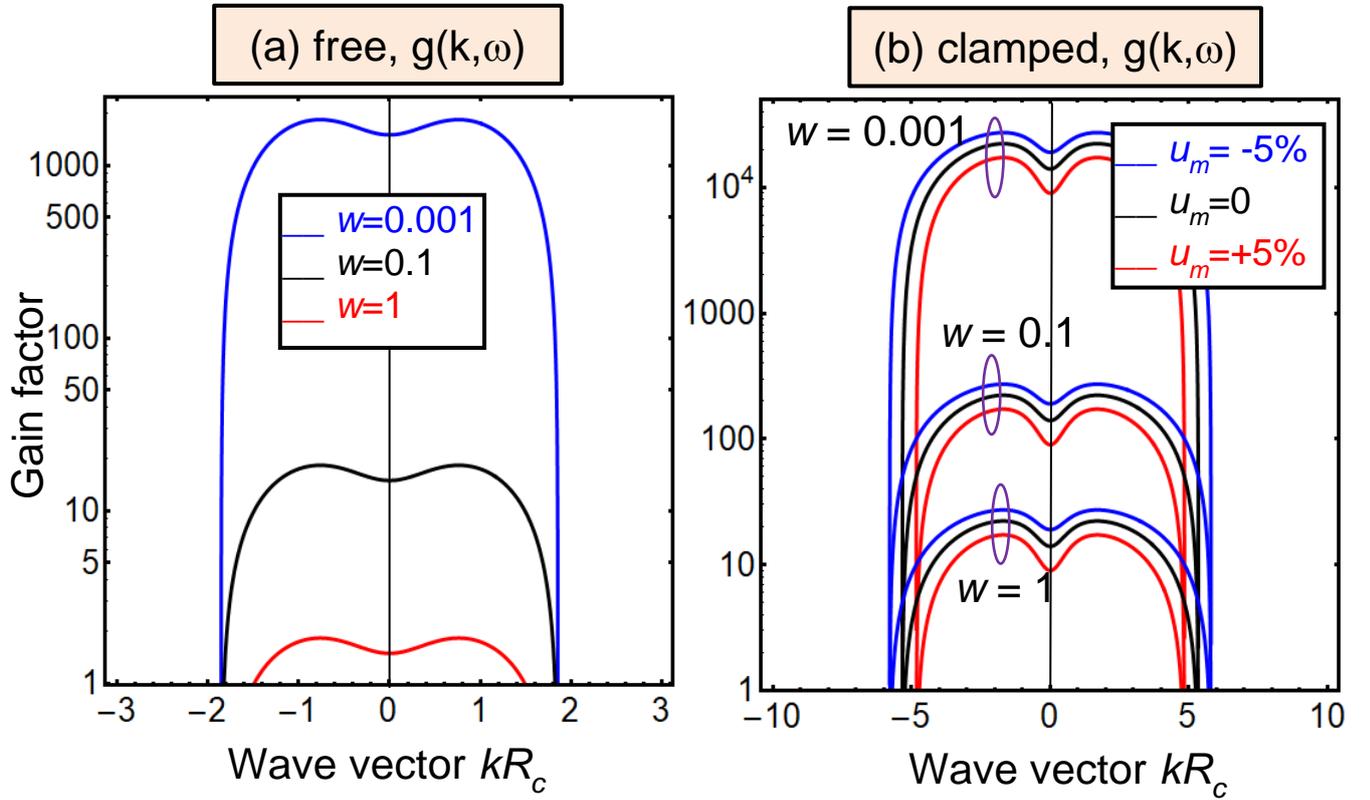

**FIGURE 5. Gain factor.** The the spectrum of gain factor, $g(\mathbf{k},\omega)$ calculated for a freestanding **(a)** and clamped **(b)** SL-TMD. Curves in the plot **(a)** are calculated for several dimensionless frequencies: $w = \omega \tau_K$, which are equal to 0.001 (blue curves), 0.1 (black curves) and 1 (red curves). Curves in the plot **(b)** are calculated for several misfit strains: $u_m = -5\%$ (blue curves), 0 (black curves) and $+5\%$ (red curves), and three frequencies $w = 0.001, 0.1$ and $1$. These three groups of curves are marked by ellipses. Other parameters are the same as in **Fig. 3**.

Finally, let us discuss the possibility to observe experimentally the predicted giant gain factors. According to [61] the gain factor for different vibrations of thymine (**Thy**) molecules on a single graphene layer is giant, $(3-5) \cdot 10^2$, meanwhile the enhancement factor for adenine (**Ade**) is $3-5$ [60]. Thy on graphene sheets shows negligible enhancement according to Ref.[59], as well for different other molecules, dyes mostly – by the way, graphene could be a simple quencher of their luminescence manifesting the RS better on a graphene SL, the enhancement factor reaches not more than 10 in most cases, and up to 47.5 in exceptional cases [64].

Thy and Ade, having similar DNA based structure, could not stem such big difference, two orders, in the enhanced spectroscopy. The reason could be in the structure of graphene layer. Really, in Ref.[61], where the big enhancement for adsorbed molecules on graphene layer was observed, the position of graphene G-mode is 1589 cm$^{-1}$ in contrast to ideal graphene layer 1580 cm$^{-1}$ [65]. G-mode at (1589-1593) cm$^{-1}$ is usually observed in single walled carbon nanotubes (curved single graphene layer) [66].

Thus, G-mode at 1589 cm$^{-1}$ in graphene indicates the numerous defects with curvature of broken graphene layers. In other cases [53, 54] with low enhancement factors for molecules adsorbed on different graphene-like and other two-dimensional structures the samples were not prepared carefully, leading to molecule crystallization, and the aggregation or initial particles without no proper sizes (unit of nm) etc.

In contrast to gold substrate, it is not easy to enhance small molecule vibrations on a graphene SL. As to enhancement of big molecules (as DNA or protein) with many cites of adsorption on the substrate, it is difficult to realize it with any type of substrate. Earlier, a big enhancement in RS was obtained for biological polymer (**polyA**) using single wall carbon nanotubes on the gold substrate with complex relief of the surface. We suppose it is due to carbon nanotube curved surface that enhanced RS signal from 1 molecule of polyA reaching giant enhancement factor about $10^5$ [67].

## 5. SUMMARY

Using LGD phenomenological approach, we have shown that a flexoelectric effect can induce the bistable out-of-plane polar displacements in rippled low-dimensional TMDs. The interesting phenomena can emerge due to the surface reconstruction resulting into its spontaneous rippling in the normal direction. Since any bending of a suspended SL can induce the strain gradient, the surface rippling immediately leads to the spontaneous polarization appearance due to the widespread flexoelectric coupling.

Another mechanism of the out-of-plane polarization appearance in ultra-thin TMDs can be the inversion symmetry breaking in the TMD completely or partially clamped to a substrate. It gives rise to a spontaneous polarization originated from the surface-induced piezoelectric effect, similarly to nanoscale ferroelectrics [21, 22]. Performed calculations proved that the out-of-plane polarization, originated from flexoelectric effect in a spontaneously rippled SL-TMDs, is bistable and reversible by a non-uniform external electric field. At the same time the spontaneous polarization, induced by a misfit strain and symmetry-sensitive surface-induced piezoelectric coupling, is electret-like and irreversible.

The special attention is paid to the analysis of the dielectric susceptibility and gain factor **k**-spectra, which enhancement is critically important for the observation of the polar phenomena in low dimensional TMDs by the surface-enhanced vibrational spectroscopy methods. We obtained that the partial mechanical clamping of a low-dimensional TMD to a substrate can essentially increase the gain factor, and the result allows making an optimistic conclusion about the feasibility of the theoretical predictions verification by SERS and SEIRA experiments.

**Acknowledgements.** Authors are very grateful to Jaekwang Lee for useful comments and stimulating discussions. This material is based upon work (S.V.K) supported by the U.S. Department of Energy (D.O.E.), Office of Science, Materials Sciences and Engineering Division, and performed at Oak Ridge National Laboratory's Center for Nanophase Materials Sciences, a U.S. DOE Office of Science User Facility. A portion of FEM was conducted at the Center for Nanophase Materials Sciences, which is a DOE Office of Science User Facility (CNMS Proposal ID: 257). This research (Y.K.) was supported by Basic Science Research Program through the National Research Foundation of Korea (NRF) funded by the Ministry of Education (No. 2019R1A6A1A03033215). A.N.M work supported by the National Academy of Sciences of Ukraine and has received funding from the European Union's Horizon 2020 research and innovation programme under the Marie Skłodowska-Curie grant agreement No 778070. G.I.D. work was supported by Ukrainian-Poland Project Development of enhanced substrates based on 2D nanomaterials for fluorescent microscopy and spectroscopy, and the project "Development of 2D materials and "smart" sensors for medical and biological purposes based thereon" of the Targeted Program of the National Academy of Sciences of Ukraine.

. 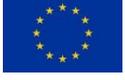

**Authors' contribution.** A.N.M. G.I.D., Y.K. and S.V.K. generated the research idea. A.N.M. provided a mathematical formulation of the problem, performed analytical calculations jointly with M.D.G., prepared illustrations and wrote the manuscript draft. E.A.E. performed numerical simulations. G.I.D., Y.K. and S.V.K. densely worked on the results interpretation and manuscript improvement.

### APPENDIX A. Derivation of Analytical Solutions

As the first step analysis we regard that the excitation field is much smaller than the coercive one, and look for analytical solutions of Eq.(3b) in the simplest one-component and one-dimensional approximation. Allowing for (6) and omitting all tensorial notations, the equation reads:

$$\Gamma \frac{\partial P}{\partial t} + \left[a + \frac{\eta}{\varepsilon_0 \varepsilon_b} + qu(x)\right] P + bP^3 - g\frac{\partial^2 P}{\partial x^2} = -f\frac{\partial u}{\partial x} + \mu \frac{\partial^2 U}{\partial t^2} + E, \qquad (A.1a)$$

where the "effective" gradient coefficient is introduced as $g = g_{12} - \frac{f^2}{c}$, and the change is caused by the flexoelectric coupling.

To obtain the analytical results, we assume that a small displacement $U(x)$ and strain $u(x) = \frac{\partial U}{\partial x}$, caused by the surface rippling or corrugation has the form of static harmonic function,

$$U(x) = U_0(\omega)\cos(k_s x + \psi), \qquad u(x) = u_m + u_0(\omega)\sin(k_s x + \psi), \qquad (A.1b)$$

where the amplitude $u_0 = -k_s U_0$. The Fourier image of the strain is $\tilde{u}(k,\omega) = \frac{u_0(\omega)}{i}\sqrt{\frac{\pi}{2}}\left[e^{i\psi}\delta(k+k_s) - e^{-i\psi}\delta(k-k_s)\right]$, where $\delta(k)$ is a Dirac-delta function.

For the strain, the polarization response in $\{k,\omega\}$-space obeys the nonlinear nonlocal equation:

$$\alpha(k,\omega)\tilde{P}(k,\omega) + \frac{qu_0}{2i}\left[e^{i\psi}\tilde{P}(k+k_s,\omega) - e^{-i\psi}\tilde{P}(k-k_s,\omega)\right] + b\widetilde{P^3}(k,\omega)=$$
$$= \tilde{E}(k,\omega) - ikf\tilde{u}(k,\omega) - \mu\omega^2\tilde{U}(k,\omega), \qquad (A.2)$$

where the $\{k,\omega\}$-spectrum of coefficient $\alpha(k,\omega)$ is renormalized by the local depolarization effects and constant components of the misfit strain

$$\alpha(k,\omega) = i\omega\Gamma + a + \frac{\eta(k,\omega)}{\varepsilon_0\varepsilon_b} + qu_m + gk^2. \qquad (A.3)$$

Since $k_s > 0$, and we are interested in a periodic linearized solution with the wavenumbers $k = \pm k_s$, the linear polarization response in $\{k, \omega\}$-space to a very small electric excitation and strain can be roughly estimated as:

$$\tilde{P}(k,\omega) \approx P_S \delta(k) + \left[\tilde{E}(k,\omega) - ikf\tilde{u}(k,\omega) - \mu\omega^2 \tilde{U}(k,\omega)\right]\chi(k,\omega). \tag{A.4a}$$

The spontaneous polarization is absent for bulk TMDs, however the irreversible polarization $P_S$ may appear in a SL-TMD due to the out-of-plane surface piezoelectric effect coupled with the misfit-induced (possibly quasi-homogeneous) local strain $u_m$ as $P_S \sim d_{13}^S u_m$. Naturally, the misfit strain is absent ($u_m = 0$) for a free-standing TMD film/flake, and so the first term in Eq.(A.4) is absent in the case.

If we would like to account for the non-locality and nonlinearity in Eq.(A.4a), the approximation for the dielectric susceptibility can be used

$$\chi(k,\omega) \approx \left(i\omega\Gamma + a + \frac{\eta(k,\omega)}{\varepsilon_0 \varepsilon_b} + qu_m - \frac{qu_0}{2i}\left[\sum_{n=0}^{\infty}\left(e^{i\psi} - (-1)^n e^{-i\psi}\right)\frac{k_s^n}{n!}\frac{\tilde{P}^{(n)}(k,\omega)}{\tilde{P}(k,\omega)}\right] + gk^2 + 3bP_S^2\right)^{-1}, \tag{A.4b}$$

The "harmonic" approximation for the derivatives $\tilde{P}^{(n)}(k,\omega) \sim (ik)^{-n}\tilde{P}(k,\omega)$ in Eq.(A.4b) yields:

$$\chi(k,\omega) \approx \left(\begin{array}{c} i\omega\Gamma + a + \frac{\eta(k,\omega)}{\varepsilon_0\varepsilon_b} + gk^2 + 3bP_S^2 + qu_m - \\ qu_0\left[\sum_{n=0}^{\infty}\frac{(-1)^{-n}}{(2n)!}\left(\frac{k_s}{k}\right)^{2n}\sin\psi + \sum_{n=1}^{\infty}\frac{(-1)^{-n}}{(2n-1)!}\left(\frac{k_s}{k}\right)^{2n-1}\cos\psi\right] \end{array}\right)^{-1}. \tag{A.4c}$$

Convergence of the series leads to

$$\chi(k,\omega) \approx \frac{1}{i\omega\Gamma + a + \frac{\eta(k,\omega)}{\varepsilon_0\varepsilon_b} + gk^2 + 3bP_S^2 + qu_m - qu_0 \sin\left[\psi - \frac{k_s}{k}\right]}. \tag{A.4d}$$

For high enough $|k| \gg |k_s|$ the series (A.4c) can be truncated, and the first terms are $\left(\sin\psi + \frac{k_s}{k}\cos\psi\right)$. Since the series in Eq.(A.4c) is phase-dependent and the phase is arbitrary, but constant, the form of physical solution should not depend on the phase Rigorously speaking, the real and imaginary parts of the susceptibility,

$$\text{Re}[\chi(k,\omega)] \approx \frac{a + \frac{\eta(k,\omega)}{\varepsilon_0\varepsilon_b} + qu_m - qu_0\sin\left[\psi - \frac{k_s}{k}\right] + gk^2 + 3bP_S^2}{(\omega\Gamma)^2 + \left(a + \frac{\eta(k,\omega)}{\varepsilon_0\varepsilon_b} + qu_m - qu_0\sin\left[\psi - \frac{k_s}{k}\right] + gk^2 + 3bP_S^2\right)^2}, \tag{A.4d}$$

$$\text{Im}[\chi(k,\omega)] \approx \frac{\omega\Gamma}{(\omega\Gamma)^2 + \left(a + \frac{\eta(k,\omega)}{\varepsilon_0\varepsilon_b} + qu_m - qu_0\sin\left[\psi - \frac{k_s}{k}\right] + gk^2 + 3bP_S^2\right)^2}, \tag{A.4e}$$

should be analyzed (or averaged) for different phases $\psi$. Limiting ourselves by an even function of the ratio $\frac{k_s}{k}$, we can put $\psi = \pm\pi/2$. After elementary calculations we obtained expressions, valid in a wider range than $|k| \gg |k_s|$:

$$\text{Re}[\chi(k,\omega)] \approx \frac{a + \frac{\eta(k,\omega)}{\varepsilon_0\varepsilon_b} + qu_m - qu_0\left(1 - \frac{k_s^2}{2(k^2 + k_s^2)}\right) + gk^2 + 3bP_S^2}{(\omega\Gamma)^2 + \left(a + \frac{\eta(k,\omega)}{\varepsilon_0\varepsilon_b} + qu_m - qu_0\left(1 - \frac{k_s^2}{2(k^2 + k_s^2)}\right) + gk^2 + 3bP_S^2\right)^2}, \tag{A.5a}$$

$$\text{Im}[\chi(k,\omega)] \approx \frac{\omega\Gamma}{(\omega\Gamma)^2 + \left(a + \frac{\eta(k,\omega)}{\varepsilon_0\varepsilon_b} + qu_m - qu_0\left(1 - \frac{k_s^2}{2(k^2 + k_s^2)}\right) + gk^2 + 3bP_S^2\right)^2}, \tag{A.5b}$$

In Eqs.(A.5) we used the Pade approximation for $\cos\left(\frac{k_s}{\sqrt{k^2+k_s^2}}\right) \approx 1 - \frac{k_s^2}{2(k^2+k_s^2)}$.

Expression (A.5) for the dielectric susceptibility in *x*-space has no physical sense, since it can be negative in some x-points. One can speculate about an "effective" permittivity only.

The condition in $\{k, \omega\}$-space

$$a + \frac{\eta(k,\omega)}{\varepsilon_0 \varepsilon_b} + qu_m - \frac{qu_0}{2} + gk^2 < 0 \qquad (A.6)$$

may indicate the potential possibility of the strain-induced transition to a ferroelectric-like phase with a robust bistable out-of-plane polarization.

Since the electric fields and strains induced by a probe can be very high, the approximate expression (8), corresponding to a linear local response, are inapplicable a priory for comparison of the proposed theoretical model with e.g. PFM experiments. One should take into account the nonlinearity and, not less important, the nonlocality of the studied polar response.

The coupled system (12), linearized with respect to polarization, acquires the form:

$$-\gamma U + \epsilon U^3 - \frac{\partial^2 U}{\partial x^2} = f \frac{\partial^2 P}{\partial x^2}, \qquad (A.8a)$$

$$P - R_c^2 \frac{\partial^2 P}{\partial x^2} \approx \chi\left(E - f \frac{\partial^2 U}{\partial x^2}\right), \qquad (A.8b)$$

where $\chi = \frac{1}{a + \frac{\eta}{\varepsilon_0 \varepsilon_b}}$ is an effective dielectric susceptibility. Using decoupling approximation for a small polarization, we omit the terms proportional to a small flexoelectric coefficient in the right-hand-side of the Eq.(A.8a), i.e. we put $\frac{\partial^2 P}{\partial x^2} \approx \chi\left(1 + \frac{R_c^2}{d^2}\right)\frac{\partial^2 E}{\partial x^2}$, and the term $f \frac{\partial^2 U}{\partial x^2}$ can be remained in the expression for polarization. Next, assuming that for a PFM probe one can use the expressions for the electric field and its second derivative at the TMD surface, $E(x) = \frac{E_0 x}{\sqrt{x^2+d^2}}$ and $\frac{\partial^2 E}{\partial x^2} \approx \frac{-E_0 x}{\sqrt{(x^2+d^2)^3}} \approx \frac{E(x)}{d^2}$ at $x^2 \ll d^2$, where *d* is the effective size of PFM tip, one can further simplify Eqs.(A.8) and rewrite them via the strain, as a parametric variable, namely

$$E[U] \approx -\frac{d^2}{f\chi\left(1+\frac{R_c^2}{d^2}\right)}\left(-\gamma U + \epsilon U^3 - \frac{\partial^2 U}{\partial x^2}\right), \qquad (A.9a)$$

$$P[U] \approx \chi\left(1 + \frac{R_c^2}{d^2}\right) E[U] - \chi f \frac{\partial^2 U}{\partial x^2}. \qquad (A.9b)$$

Equations (15) in the main text follow from Eqs.(A.9) in a s straightforward way.

## REFERENCES

[1] J. F. Scott, Applications of Modern Ferroelectrics, Science **315**, 954 (2007)

[2] Ferroelectric thin films: Review of materials, properties, and applications, N. Setter and D. Damjanovic..., J. Appl. Phys., **100**, 051606 (2006)

[3] A.K. Tagantsev, Piezoelectricity and flexoelectricity in crystalline dielectrics, Phys. Rev B, **34**, 5883 (1986).

[4] M. S. Majdoub, P. Sharma, and T. Cagin, Title, Phys. Rev B **77**, 125424 (2008).

[5] "Flexoelectricity in Solids: From Theory to Applications". Ed. by A.K. Tagantsev and P.V. Yudin, World Scientific (2016)

[6] B. Wang, Y. Gu, S. Zhang, L.-Q. Chen. Flexoelectricity in solids: Progress, challenges, and perspectives. Progress in Materials Science, **106**, 100570 (2019)

[7] Kim, Sung Kyun, Ravi Bhatia, Tae-Ho Kim, Daehee Seol, Jung Ho Kim, Hyun Kim, Wanchul Seung, Yunseok Kim, Young Hee Lee, and Sang-Woo Kim. "Directional dependent piezoelectric effect in CVD grown monolayer MoS2 for flexible piezoelectric nanogenerators." Nano Energy 22, 483-489 (2016)

[8] Michael N. Blonsky, Houlong L. Zhuang, Arunima K. Singh, and Richard G. Hennig. "Ab initio prediction of piezoelectricity in two-dimensional materials." ACS Nano 9, 9885-9891 (2015).

[9] M. Menderes Alyörük, Yierpan Aierken, Deniz Çakır, Francois M. Peeters, and Cem Sevik. "Promising piezoelectric performance of single layer transition-metal dichalcogenides and dioxides." J. Phys. Chem. C 119, 23231 (2015).

[10] Wenzhuo Wu, Lei Wang, Yilei Li, Fan Zhang, Long Lin, Simiao Niu, Daniel Chenet et al. "Piezoelectricity of single-atomic-layer MoS 2 for energy conversion and piezotronics." Nature 514, 470–474 (2014).

[11] S.N. Shirodkar and U.V. Waghmare, Emergence of ferroelectricity at a metal-semiconductor transition in a 1 T monolayer of $MoS_2$, Phys. Rev. Lett. 112, 157601 (2014).

[12] Shuoguo Yuan, Xin Luo, Hung Lit Chan, Chengcheng Xiao, Yawei Dai, Maohai Xie and Jianhua Hao, Room-temperature ferroelectricity in MoTe2 down to the atomic monolayer limit, Nature Communications 10, 1775 (2019).

[13] P. Johari, and V. B. Shenoy. Tuning the electronic properties of semiconducting transition metal dichalcogenides by applying mechanical strains. ACS nano 6, 5449-5456 (2012).

[14] K.-A. N. Duerloo, Y. Li, and E. J. Reed, Structural phase transitions in two-dimensional Mo- and W-dichalcogenide monolayers, Nat. Commun. 5, 4214 (2014).

[15] A.N. Enyashin, L. Yadgarov, L. Houben, I. Popov, M. Weidenbach, R. Tenne, M. Bar-Sadan, and G. Seifert. New route for stabilization of 1T-WS2 and MoS2 phases. J. Phys. Chem. C 115, 24586–24591 (2011).

[16] G. Eda, T. Fujita, H. Yamaguchi, D. Voiry, D. Chen, M. Chhowalla, Coherent Atomic and Electronic Heterostructures of Single-Layer MoS2. ACS Nano 6, 7311–7317 (2012).

[17] M. Wu, S. Dong, K. Yao, J. Liu, and X. C. Zeng. Ferroelectricity in covalently functionalized two-dimensional materials: integration of high-mobility semiconductors and nonvolatile memory. Nano Lett. 16, 7309-7315 (2016).

[18] L. Li, and M. Wu. Binary compound bilayer and multilayer with vertical polarizations: two-dimensional ferroelectrics, multiferroics, and nanogenerators. ACS nano 11, 6382-6388 (2017).

[19] S. Yuan, X. Luo, H.L. Chan, C. Xiao, Y. Dai, M. Xie, and J. Hao. Room-temperature ferroelectricity in MoTe 2 down to the atomic monolayer limit. Nat. Commun. 10, 1775 (2019).

[20] J.H. Choi, S.H. Jhi. Origin of robust out-of-plane ferroelectricity in d1T-MoS 2 monolayer. J.Phys.: Cond. Matter 32 045702 (2020).

[21] M.D. Glinchuk, A.N. Morozovska, E.A. Eliseev. Ferroelectric thin films phase diagrams with self-polarized phase and electret state. J. Appl. Phys. 99, 114102 (2006).

[22] M.D. Glinchuk, A.N. Morozovska. The internal electric field originating from the mismatch effect and its influence on ferroelectric thin film properties. J. Phys.: Condens. Matter Vol. 16, 3517–3531 (2004).

[23] Pankaj, Sharma, Fei-Xiang Xiang, Ding-Fu Shao, Dawei Zhang, Evgeny Y. Tsymbal, Alex R. Hamilton, and Jan Seidel. "A room-temperature ferroelectric semimetal." Sci. Adv. 5, eaax5080 (2019).

[24] Zaiyao Fei, Wenjin Zhao, Tauno A. Palomaki, Bosong Sun, Moira K. Miller, Zhiying Zhao, Jiaqiang Yan, Xiaodong Xu, and David H. Cobden. Ferroelectric switching of a two-dimensional metal. Nature 560, 336 (2018).

[25] Alessio Filippetti, Vincenzo Fiorentini, Francesco Ricci, Pietro Delugas, and Jorge Íñiguez. "Prediction of a native ferroelectric metal." Nature Communications 7, 11211 (2016).

[26] Jannik C. Meyer, A. K. Geim, M. I. Katsnelson, K. S. Novoselov, T. J. Booth and S. Roth, The structure of suspended graphene sheets, Nature 446, 60 (2007).

[27] T. Dumitrica, C. M. Landis, B. I. Yakobson, Curvature-induced polarization in carbon nanoshells, Chemical Physics Letters, 360, 182 (2002)

[28] S. V. Kalinin, and V. Meunier. Electronic flexoelectricity in low-dimensional systems. Phys. Rev. B, 77, 033403 (2008).

[29] Seunghun Kang, Sera Jeon, Sera Kim, Daehee Seol, Heejun Yang, Jaekwang Lee, and Yunseok Kim. "Tunable out-of-plane piezoelectricity in thin-layered MoTe2 by surface corrugation-mediated flexoelectricity." ACS Appl. Mater. Inter. 10, 27424 (2018).

[30] Ivan Naumov, Alexander M. Bratkovsky, and V. Ranjan. "Unusual Flexoelectric Effect in Two-Dimensional Noncentrosymmetric s p 2-Bonded Crystals." Phys. Rev. Lett. 102, 217601 (2009).

[31] W. Shi, Y, Guo, Z. Zhang, and W. Guo. J. Phys. Chem. Lett. 9, 6841 (2018).

[32] P V Yudin and A K Tagantsev. Fundamentals of flexoelectricity in solids. Nanotechnology, 24, 432001 (2013).

[33] . V.S. Mashkevich, and K.B. Tolpygo, *Zh.Eksp.Teor.Fiz.* 31, 520 (1957) [*Sov.Phys. JETP*, 4, 455 (1957)].

[34] Sh. M. Kogan, " Piezoelectric effect under an inhomogeneous strain and an acoustic scattering of carriers of current in crystals "Solid State Physics, Vol. 5, 10, 2829 (1963)

[35] A. Kvasov, and A. K. Tagantsev, Dynamic flexoelectric effect in perovskites from first-principles calculations, Phys. Rev. B 92, 054104 (2015).

[36] K. T. Kang, J. Park, D. Suh, and W. S. Choi. Advanced Materials, 31 (34), 1803732 (2019)

[37] A.N. Morozovska, E.A. Eliseev, G.S. Svechnikov, and S.V. Kalinin. Nanoscale electromechanics of paraelectric materials with mobile charges: Size effects and nonlinearity of electromechanical response of $SrTiO_3$ films. Phys. Rev. B 84, 045402 (2011).

[38] Y. Kim, A. N. Morozovska, A. Kumar, S. Jesse, E. A. Eliseev, F. Alibart, D. Strukov, and S. V. Kalinin. Ionically-Mediated Electromechanical Hysteresis in Transition Metal Oxides. ACS Nano 6, (8), 7026–7033 (2012)

[39] Michael Hoffmann, Franz PG Fengler, Melanie Herzig, Terence Mittmann, Benjamin Max, Uwe Schroeder, Raluca Negrea, Pintilie Lucian, Stefan Slesazeck, and Thomas Mikolajick. "Unveiling the double-well energy landscape in a ferroelectric layer." Nature 565, 464 (2019).

[40] M. D. Glinchuk, A. N. Morozovska, A. Lukowiak, W. Stręk, M. V.Silibin, D. V.Karpinsky, Y. Kim and S. V. Kalinin. Possible Electrochemical Origin of Ferroelectricity in $HfO_2$ Thin Films. Journal of Alloys and Compounds, 830, 153628 (2020)

[41] J. Berry, S. Zhou, J. Han, D. J. Srolovitz, and M. P. Haataja. Dynamic phase engineering of bendable transition metal dichalcogenide monolayers. Nano letters 17, 2473-2481 (2017).

[42] J. Berry, S. Zhou, J. Han, D. J. Srolovitz, and M. P. Haataja. Domain morphology and mechanics of the H/T′ transition metal dichalcogenide monolayers. Phys. Rev. Mater. 2, 114002 (2018).

[43] A. N. Morozovska, E. A. Eliseev, K. D. Stubbs, R. Vasudevan, Y. Kim, and S. V. Kalinin. Phase Diagrams of Single Layer Two-Dimensional Transition Metal Dichalcogenides: Landau Theory. Phys. Rev. B, **101**, 195424 (2020)

[44] E.A. Eliseev, A.N. Morozovska, M.D. Glinchuk, and R. Blinc. Spontaneous flexoelectric/flexomagnetic effect in nanoferroics. Phys. Rev. B. **79**, 165433-1-10 (2009).

[45] A.N. Morozovska, E.A. Eliseev, A.K. Tagantsev, S.L. Bravina, Long-Qing Chen, and S.V. Kalinin. Thermodynamics of electromechanically coupled mixed ionic-electronic conductors: Deformation potential, Vegard strains, and flexoelectric effect. Phys. Rev. **B 83**, 195313 (2011).

[46] E. A. Eliseev, I. S. Vorotiahin, Y. M. Fomichov, M. D. Glinchuk, S. V. Kalinin, Y. A. Genenko, and A. N. Morozovska. Defect driven flexo-chemical coupling in thin ferroelectric films. Physical Review **B, 97,** 024102 (2018)

[47] L.D. Landau, and E. M. Lifshitz. *Course of Theoretical Physics. Vol. 8: Electrodynamics of Continous Media*. Oxford, 1963.

[48] P. V. Yudin, R. Ahluwalia, A. K. Tagantsev. Upper bounds for flexocoupling coefficients in ferroelectrics, Appl.Phys.Lett. **104**(8), 082913 (2014)

[49] A. N. Morozovska, E. A. Eliseev, C. M. Scherbakov, and Y. M. Vysochanskii, The influence of elastic strain gradient on the upper limit of flexocoupling strength, spatially-modulated phases and soft phonon dispersion in ferroics. *Phys. Rev.* B 94, 174112 (2016)

[50] E. Gnecco, P. Pedraz, P. Nita, F. Dinelli, S. Napolitano and P. Pingue. Surface rippling induced by periodic instabilities on a polymer surface. New J. Phys.17, 032001 (2015)

[51] W. Ma and L.E. Cross. Flexoelectricity of barium titanate. Appl. Phys. Lett., **88**, 232902 (2006).

[52] A. K. Saha, and S. K Gupta. Multi-Domain Negative Capacitance Effects in Metal-Ferroelectric-Insulator-Semiconductor/Metal Stacks: A Phase-field Simulation Based Study. Accepted to Scientific Reports (2020)

[53] M.D. Glinchuk, A.N. Morozovska & E.A. Eliseev. Correlation Radius in Thin Ferroelectric Films. Ferroelectrics 400, No 1, 243-254 (2010).

[54] A.N. Morozovska, E.A. Eliseev. Surface and finite size effect on fluctuations dynamics in nanoparticles with long-range order. J. Appl. Phys. 107, 044101 (2010).

[55] S.V. Kalinin, B.J. Rodriguez, S. Jesse, E. Karapetian, B. Mirman, E.A. Eliseev, A.N. Morozovska. Nanoscale Electromechanics of Ferroelectric and Biological Systems: A New Dimension in Scanning Probe Microscopy (review). Annual Review of Materials Research 37, 189–238 (2007).

[56] A.N. Morozovska, E.A. Eliseev, S.L. Bravina, S.V. Kalinin. Resolution Function Theory in Piezoresponse Force Microscopy: Domain Wall Profile, Spatial Resolution, and Tip Calibration. Phys. Rev. B 75, 174109 (2007).

[57] S.V. Kalinin, S. Jesse, B.J. Rodriguez, Y.H. Chu, R. Ramesh, E.A. Eliseev, A.N. Morozovska // Probing the role of single defects on thermodynamics of electric-field induced phase transitions / Phys. Rev. Lett. 100, № 15, 155703-1-4 (2008).

[58] K. P. Kelley, Yao Ren, A. N. Morozovska, E. A. Eliseev, Y. Ehara, H. Funakubo, T. Giamarchi, N. Balke, R. K. Vasudevan, Ye Cao, S. Jesse, S. V. Kalinin. Dynamic manipulation in piezoresponse force microscopy: creating non-equilibrium phases with large electromechanical response. (reference number: NM20010092), (https://arxiv.org/abs/2001.03586)

[59] G. Dovbeshko, O. Gnatyuk, O. Fesenko, A. Rynder, and O. Posudievsky. Enhancement of infrared absorption of biomolecules absorbed on single-wall carbon nanotubes and graphene nanosheets, Journal of Nanophotonics,6(1),061711, 2012, DOI: 10.1117/1.JNP.6.061711

[60] L. Dolgov, D. Pidgirnyi, G. Dovbeshko, T. Lebedieva, V. Kiisk, S. Heinsalu, S. Lange, R. Jaaniso, I. Sildos: Graphene enhanced Raman scattering from the adenine molecules. Nanoscale Research Letters 11:197, 2016, DOI 10.1186/s11671-016-1418-5

[61] O. Fesenko, G. Dovbeshko, A. Dementjev, Re. Karpicz, T. Kaplas and Y. Svirko. Graphene-enhanced Raman spectroscopy of thymine adsorbed on single-layer graphene. Nanoscale Res Lett. **10,** 163 (2015). https://doi.org/10.1186/s11671-015-0869-4

[62] G. Dovbeshko, O. Fesenko, A. Dementjev, R. Karpicz, V. Fedorov and O. Yu Posudievsky, Coherent anti-Stokes Raman scattering enhancement of thymine adsorbed on graphene oxide. *Nanoscale Res Lett* **9,** 263 (2014).

[63] V. Kosobukin, "The effect of enhancement the external field near the surface of metal and its manifestation in spectroscopy," Surface Phys. Chem. Mech. 12, 5–20 (1983)

[64] Huang, S., Ling, X., Liang, L., Song, Y., Fang, W., Zhang, J., et al, Dresselhaus, M. S. Molecular Selectivity of Graphene-Enhanced Raman Scattering. Nano Letters, 15(5), 2892–2901. (2015). doi:10.1021/nl5045988

[65] Ferrari, A. C., Meyer, J. C., Scardaci, V., Casiraghi, C., Lazzeri, M., Mauri, F., … Geim, A. K. Raman Spectrum of Graphene and Graphene Layers. Physical Review Letters, **97**(18) (2006). doi:10.1103/physrevlett.97.187401

[66] G. I. Dovbeshko , O. P. Gnatyuk , A. N. Nazarova , Yu. I.Sementsov & E. D. Obraztsova (2005): Vibrational Spectra of Carbonaceous Materials: A SEIRA Spectroscopy versus FTIR and Raman, Fullerenes, Nanotubes and Carbon Nanostructures, 13:S1, 393-400 http://dx.doi.org/10.1081/FST-200039387

[67] Galina Dovbeshko, O. Fesenko, O. Gnatyk, Y. Shtogun, L. Woods, S. Bertarione, A. Damin, D. Scarano, Z. Adriano. Nucleic acid interaction and interfaces with single-walled carbon nanotubes. Carbon nanotubes (Edited by: Jose Mauricio Marulanda) – India, March 2010. P 697-720.